\documentclass[12pt]{article}
\usepackage{ssi}
\usepackage{times}
\input{psfig}

\begin{document}

\def\fun#1#2{\lower3.6pt\vbox{\baselineskip0pt\lineskip.9pt
  \ialign{$\mathsurround=0pt#1\hfil##\hfil$\crcr#2\crcr\sim\crcr}}}
\def\lesssim{\mathrel{\mathpalette\fun <}}
\def\gtrsim{\mathrel{\mathpalette\fun >}}

\long\def\comment#1{}

\def\VEV#1{\left\langle #1\right\rangle}
\def\sec{\ifmmode \,\, {\rm sec} \else sec \fi}
\def\eV {\ifmmode \,\, {\rm eV} \else eV \fi}
\def\keV{\ifmmode \,\, {\rm keV} \else keV \fi}
\def\MeV{\ifmmode \,\, {\rm MeV} \else MeV \fi}
\def\GeV{\ifmmode \,\, {\rm GeV} \else GeV \fi}
\def\TeV{\ifmmode \,\, {\rm TeV} \else TeV \fi}
\def\fm{\ifmmode \,\, {\rm fm} \else TeV \fi}
\def\pbarn{\ifmmode \,\, {\rm pb} \else pb \fi}
\def\km{\ifmmode {\rm km}\, \else km \fi}
\def\Mpc{\ifmmode {\rm Mpc}\, \else Mpc \fi}
\def\Gyr{\ifmmode {\rm Gyr}\, \else Gyr \fi}
\def\Mx{{m_{\chi}}}
\def\Mq{m_q}
\def\Msq{m_{\tilde q}}
\def\ra{\rightarrow}
\def\fun#1#2{\lower3.6pt\vbox{\baselineskip0pt\lineskip.9pt
  \ialign{$\mathsurround=0pt#1\hfil##\hfil$\crcr#2\crcr\sim\crcr}}}
\def\la{\mathrel{\mathpalette\fun <}}
\def\ga{\mathrel{\mathpalette\fun >}}
\def\order{{\cal O}}
\def\etal{{\rm et al.}}
\def\neut{{\tilde\chi}}
\def\mx{{m_{\chi}}}
\def\tanb{\tan\beta}
\def\Msf{ m_{\tilde f}}
\def\sbar#1{\kern 0.8pt
        \overline{\kern -0.8pt #1 \kern -0.8pt}
        \kern 0.8pt}  
\def\Nzsq{\VEV{Nz^2}}
\def\meter{\ifmmode \,\, {\rm m} \else m \fi}
\def\yr {\ifmmode \,\, {\rm yr} \else yr \fi}
\def\Ein{{E_{\rm in}}}
\def\sr{\ifmmode \,\, {\rm sr} \else sr \fi}
\def\sigann{(\sigma_A v)_{26}}
\def\kmsec{km sec$^{-1}$}
\def\Rf{\baselineskip=12pt\parindent=0pt \hangindent=3pc \hangafter=1}
\def\minim{{\rm min}}
\def\Msolar{M_\odot}
\def\hatn{{\bf \hat n}}

\def\slashchar#1{\setbox0=\hbox{$#1$}           
   \dimen0=\wd0                                 
   \setbox1=\hbox{/} \dimen1=\wd1               
   \ifdim\dimen0>\dimen1                        
      \rlap{\hbox to \dimen0{\hfil/\hfil}}      
      #1                                        
   \else					
      \rlap{\hbox to \dimen1{\hfil$#1$\hfil}}   
      /                                         
   \fi}

\title{New Views of Cosmology and the Microworld}

\author{Marc Kamionkowski\thanks{Supported by NASA NAG5-9821 and DoE
DE-FG03-92-ER40701.
\vskip 0.5in 
\noindent
\copyright\ 2002 by Marc Kamionkowski. }\\ 
California Institute of Technology \\
Mail Code 130-33 \\
Pasadena, CA 91125}

\maketitle

\begin{abstract}
\baselineskip=16pt
The past few years have seen several breakthroughs
in particle astrophysics and cosmology.  In several cases, new
observations can only be explained with the introduction of new
fundamental physics.  In this talk I summarize some of these
recent advances and describe several areas where progress may
well be made in the future.  More specifically, I focus on
supersymmetric and axion dark matter, self-interacting dark
matter, cosmic-microwave-background and large-scale-structure
tests of inflation, and the dark-energy problem.
\end{abstract}

\section{Introduction}

The aim of particle physics is to understand the fundamental
laws of nature.  The primary tools in this endeavor have been and
continue to be accelerator experiments which provide
controlled environments for precise experiments.  However, this
avenue may have limitations, especially considering that many of
the most promising ideas for new physics beyond the standard
model---e.g., grand unification and quantum gravity---can be
tested only at energies many orders of magnitude greater than
those accessible by current and planned accelerators.

We have recently seen an increased effort to develop
cosmological and astrophysical tools to search for and/or
constrain new physics beyond the standard model.  These efforts
do have precedents: Newton's law
of universal gravitation was motivated by planetary orbits.
Helium was first discovered in the solar spectrum.  Positrons
and muons were both discovered in cosmic-ray experiments.  For
several years before the advent of LEP, big-bang nucleosynthesis
provided the only bound to the number of light neutrinos
(e.g., Ref.~\citenum{schramm}).  Cosmology for a long time
provided by far the most stringent upper limits to
stable-neutrino masses.

The promise of using astrophysics to
study fundamental physics has also been realized more recently,
most notably with the evidence for inflation from the cosmic
microwave background (CMB), evidence for an accelerated
cosmological expansion, and atmospheric and solar neutrinos.

At the Snowmass 2001 workshop on the future of high-energy
physics, a working group (P4) was convened to identify
opportunities for advances at the interface of particle physics,
astrophysics, and cosmology.\cite{p4}
This working group covered a broad range of topics,
subdivided into eight topical groups:  (1) dark matter and relic
particles; (2) gamma rays and X-rays; (3) the CMB
and inflation; (4) structure formation and
cosmological parameters; (5) cosmic rays; (6) gravitational
radiation; (7) neutrino astrophysics; and (8) the early Universe and
tests of fundamental physics.  Recent advances in
these areas include the following:

(1) CMB measurements have now mapped the location of
the first acoustic peak in the CMB power spectrum, which
determines the geometry,\cite{KamSpeSug94} and found that the
total energy density of the Universe (in units of the critical
density) is $\Omega_{\rm tot}=1.00^{+0.03}_{-0.02}$, providing for the very
first time strong evidence that of the three possibilities
(open, closed, or flat), the spatial geometry of the Universe is
flat.\cite{Miletal99,deBetal00,Hanetal00,dasiT,Masetal02,archeops}
These CMB experiments moreover support the hypothesis that
large-scale structure grew from primordial density fluctuations
that look much like those predicted by inflation.

(2) The discrepancy between a matter density
$\Omega_m \simeq 0.3$ and $\Omega_{\rm tot} \simeq 1$
provides independent corroboration of the remarkable recent
supernova evidence\cite{Peretal99,Rieetal98} that suggests that
$\sim70\%$ of the energy density of the Universe is in the form
of some mysterious and theoretically unanticipated
negative-pressure ``dark energy''.

(3) CMB data verify, through a completely independent avenue,
the big-bang nucleosynthesis prediction that baryons make up only
$\sim$5\% of the critical density.  When combined with 
dynamical and CMB evidence for a nonrelativistic-matter
density of 30\% of critical, we infer that 25\% of the total
density of the Universe must be in the form of nonbaryonic dark
matter (the best bet being supersymmetric particles or
axions).

(4) The sensitivities of experiments to directly detect
supersymmetric and axion dark matter have been improved by
several orders of magnitude and are now probing the
cosmologically-relevant regions of parameter space.

(5) Neutrinos from astrophysical sources (atmospheric and
solar neutrinos) have provided convincing evidence for
neutrino oscillations and thus demonstrate that very concrete
advances in fundamental physics can occur with astrophysical
sources.

Here I summarize a few of the topics of the Snowmass P4
working group, focusing on several subjects that I find
particularly interesting and providing updates in several cases
where there has been progress during the past year. 

\section{Particle Dark Matter}

Almost all astronomers will agree that most of the 
mass in the Universe is nonluminous.  Dynamics of clusters of
galaxies have long suggested a universal nonrelativistic-matter
density $\Omega_m\simeq0.1-0.3$  (in units of the critical
density).  It has also been appreciated for a
long time that if there were no matter beyond the luminous matter we see,
the duration of the epoch of structure formation would be
very short, thereby requiring fluctuations in the CMB
considerably larger than those observed.\cite{KamSpe94}

\begin{figure}[htbp]
\centerline{\psfig{file=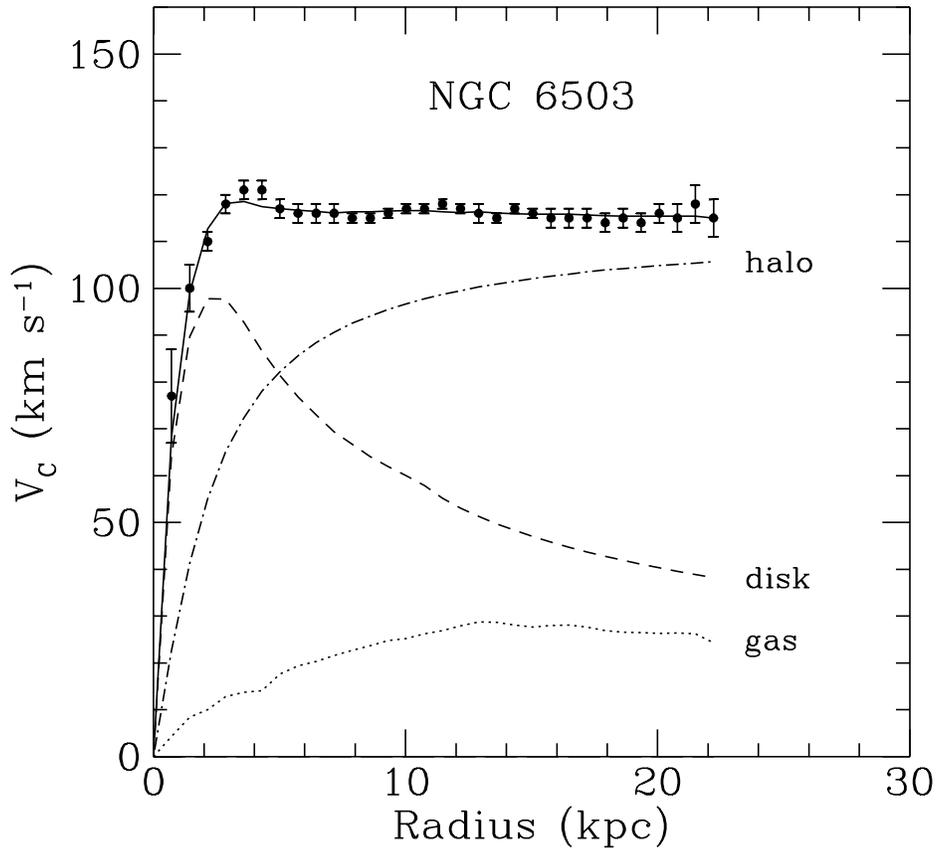,width=5in}}
\caption{Rotation curve for the spiral galaxy NGC6503.  The points
	 are the measured circular rotation velocities as a
	 function of distance from the center of the galaxy.
	 The dashed and dotted curves are the contribution to
	 the rotational velocity due to the observed disk and
	 gas, respectively, and the dot-dash curve is the
	 contribution from the dark halo.  From
	 Ref.~\protect\citenum{broeils}.}
\label{rotationfigure}
\end{figure}

However, the most robust observational evidence for the
existence of dark matter has always involved galactic dynamics.
There is simply not enough luminous matter observed in spiral
galaxies to account for their observed rotation curves (for
example, that for NGC6503 shown in Fig.~\ref{rotationfigure};
from Ref.~\citenum{broeils}).  These rotation curves imply the
existence of a
diffuse halo of dark matter that vastly outweighs and extends
much further than the luminous component.  Summing the
contributions from all galaxies, we infer that dark matter
associated with galaxies contributes $\Omega_{\rm halo} \ga
0.1$.  On the other hand, big-bang nucleosynthesis suggests
a baryon density $\Omega_b\la0.1$ (Ref.~\citenum{bbn}).
Thus, the bulk of the halo must be nonbaryonic.
In the past few years, the existence of nonbaryonic dark
matter has received independent and precise confirmation with
new CMB results alluded to above.  There is simply no good fit
to the CMB power spectrum without nonbaryonic dark matter.  The
data require a  nonbaryonic-dark-matter density $\Omega_{\rm dm}
h^2=0.13\pm0.04$  (and $h$ is the Hubble parameter in units of 100
km~sec$^{-1}$~Mpc$^{-1}$).

So, what could this dark matter be?  A neutrino species of mass
${\cal O}(10\, {\rm eV})$ could provide the right dark-matter
density, but N-body simulations of structure formation in a
neutrino-dominated Universe do a poor job of reproducing the
observed structure.\cite{Nbody}  Furthermore, it is difficult
to see (essentially the Pauli principle) how such a neutrino
could make up the halo dark matter.\cite{gunn} It appears
likely then that some exotic particle dark matter is required.

For the past two decades, the two leading candidates from
particle theory have been weakly-interacting massive particles
(WIMPs), such as the lightest superpartner (LSP) in
supersymmetric extensions of the standard
model,\cite{jkg,bergstrom} and axions.\cite{axion}

\subsection{Weakly-Interacting Massive Particles}

Suppose that in addition to the known particles of the
standard model, there exists a new stable weakly-interacting
massive particle (WIMP), $\chi$.  At sufficiently early times
after the big bang, when the temperatures are
greater than the mass of the particle, $T\gg m_\chi$, the
equilibrium number density of such particles is $n_\chi \propto
T^3$, but for lower temperatures, $T\ll m_\chi$, the equilibrium
abundance is exponentially suppressed, $n_\chi \propto
e^{-m_\chi/T}$.  If the expansion of the Universe were slow enough
that  thermal equilibrium were always maintained, the number of
WIMPs today would be infinitesimal.  However, the Universe is
not static, so equilibrium thermodynamics is not the entire story.

%
\begin{figure}[htbp]
\centerline{\psfig{file=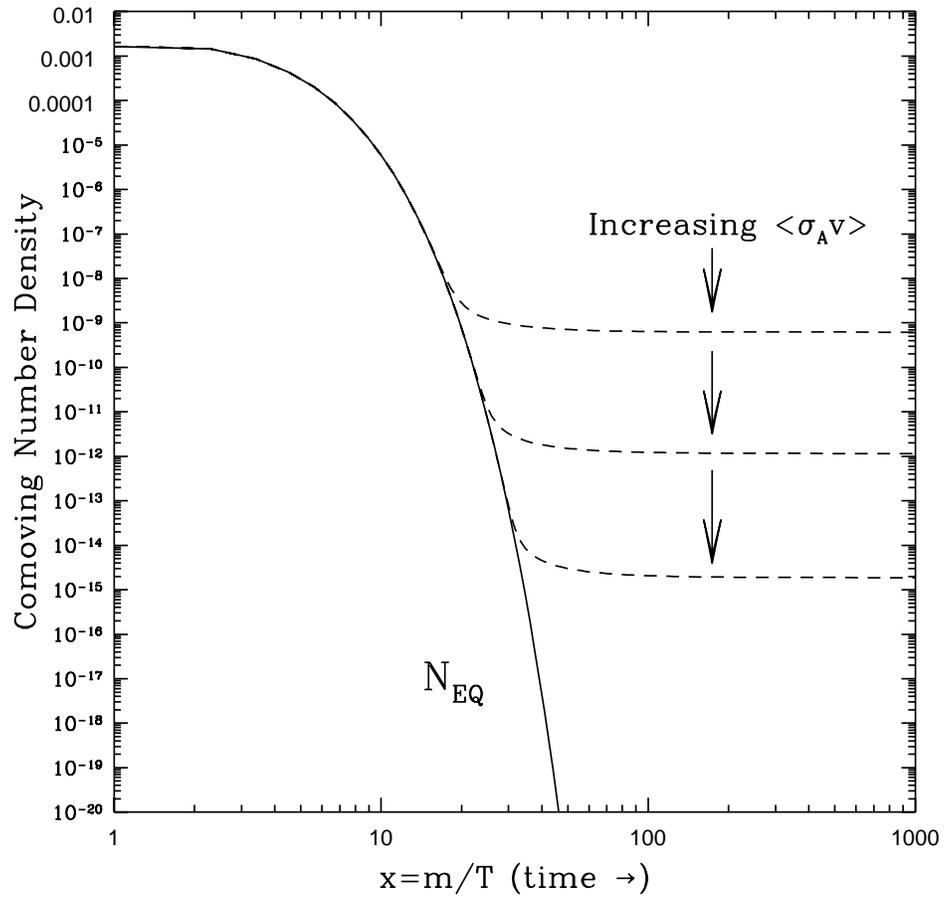,width=5in}}
\caption{Comoving number density of WIMPs in the early
	Universe.  The dashed curves are the actual abundances
	for different annihilation cross sections,
	and the solid curve is the equilibrium abundance.}
\label{YYY}
\end{figure}

At high temperatures ($T\gg m_\chi$), $\chi$'s are abundant and
rapidly converting to lighter particles and {\it vice versa}
($\chi\bar\chi\leftrightarrow l\bar l$, where $l\bar l$ are quark-antiquark and
lepton-antilepton pairs, and if $m_\chi$ is greater than the mass of the
gauge and/or Higgs bosons, $l\bar l$ could be gauge- and/or Higgs-boson
pairs as well).  Shortly after $T$ drops below $m_\chi$ the number
density of $\chi$'s drops exponentially, and the rate for annihilation of
$\chi$'s, $\Gamma=\VEV{\sigma v} n_\chi$---where $\VEV{\sigma v}$ is the
thermally averaged total cross section $\sigma$ for annihilation
of $\chi\bar\chi$ into lighter particles times relative velocity
$v$---drops below the expansion rate, $\Gamma\la H$.  At this
point, the $\chi$'s cease to annihilate efficiently, they fall
out of equilibrium, and a relic cosmological abundance remains.
The equilibrium (solid curve) and actual (dashed
curve) abundances of WIMPs per comoving volume are plotted in
Fig. \ref{YYY} as a function of $x\equiv m_\chi/T$ (which
increases with increasing time).  As the annihilation cross
section is increased the WIMPs stay in equilibrium longer, so we
are left with a smaller relic abundance when they do finally
freeze out.
An approximate solution to the Boltzmann equation yields the
cosmological WIMP abundance (in units of the critical density
$\rho_c$),
\begin{equation}
     \Omega_\chi ={m_\chi n_\chi \over \rho_c h^2}\simeq
     \left({3\times 10^{-27}\,{\rm cm}^3 \, {\rm sec}^{-1} \over
     \langle \sigma_A v\rangle }\right) h^{-2}.
\label{eq:abundance}
\end{equation}
The result is to a first approximation independent of the WIMP
mass and is fixed primarily by the annihilation cross section.

The WIMP velocities at freeze-out are typically some appreciable
fraction of the speed of light.  Therefore, from
Eq.~(\ref{eq:abundance}), the WIMP will have a cosmological
abundance of order unity today if the annihilation cross section
is roughly $10^{-9}$ GeV$^{-2}$.  Curiously, this is the order
of magnitude one would expect from a typical electroweak cross
section, 
\begin{equation}
     \sigma_{\rm weak} \simeq {\alpha^2 \over m_{\rm weak}^2},
\end{equation}
where $\alpha \simeq {\cal O}(0.01)$ and $m_{\rm weak} \simeq
{\cal O}(100\, {\rm GeV})$.  The numerical constant in
Eq.~(\ref{eq:abundance}) needed to provide $\Omega_\chi\sim1$
comes essentially from the age of the Universe.  But why should
the age of the Universe have anything to do with
the age of the Universe?  This unanticipated coincidence suggests that if a
new, as yet undiscovered, stable massive particle with electroweak
interactions exists, then it should have a relic density of
order unity and is therefore a natural dark-matter
candidate.  This has been the argument driving the massive
experimental effort to detect WIMPs.

The first WIMPs considered were massive Dirac or Majorana
neutrinos with masses in the range of a few GeV to a few TeV.
(Due to the Yukawa coupling which gives a neutrino its mass, the
neutrino interactions become strong above a few TeV, and it no
longer remains a suitable WIMP candidate.\cite{unitarity})  LEP ruled out
neutrino masses below half the $Z^0$ mass.  Furthermore, heavier
Dirac neutrinos have been ruled out as the primary component of
the Galactic halo by direct-detection experiments (described
below),\cite{heidelberg} and heavier Majorana neutrinos have
been ruled out by indirect-detection
experiments\cite{kamiokande,imb,baksan,macronew,superK,AMANDA}
(also described below) over much
of their mass range.  Therefore, Dirac neutrinos cannot comprise
the halo dark matter;\cite{griestsilk} Majorana neutrinos can,
but only over a small range of fairly large masses.  

A much more promising WIMP candidate comes from
electroweak-scale supersymmetry
(SUSY).\cite{jkg,bergstrom,haberkane}  SUSY was
hypothesized in particle physics to cure the naturalness problem
with fundamental Higgs bosons at the electroweak scale.
Coupling-constant unification at the GUT scale seems to be
improved with SUSY, and SUSY is an essential ingredient
in theories that unify gravity with the other three fundamental
forces.

The existence of a new symmetry,
$R$-parity, in SUSY theories guarantees that the lightest
supersymmetric particle (LSP) is stable.
In the minimal supersymmetric extension of the
standard model (MSSM), the LSP is usually the neutralino, a linear
combination of the supersymmetric partners of the photon, $Z^0$,
and Higgs bosons.  Another possibility are sneutrinos, but
these particles interact like neutrinos and have been ruled out
over most of the available mass range.\cite{sneutrino}
Given a SUSY model, the cross section for
neutralino annihilation to lighter particles, and thus the relic
density, can be calculated.  The
mass scale of supersymmetry must be of order the weak scale to
cure the naturalness problem, and the neutralino will have only
electroweak interactions.  Therefore, it is to be expected that
the cosmological neutralino abundance is of order unity.  In
fact, with detailed calculations, one finds that the neutralino
abundance in a very broad class of supersymmetric extensions of
the standard model is near unity and can therefore account for
the dark matter in our halo.\cite{ellishag}

\begin{figure}[htbp]
\centerline{\psfig{file=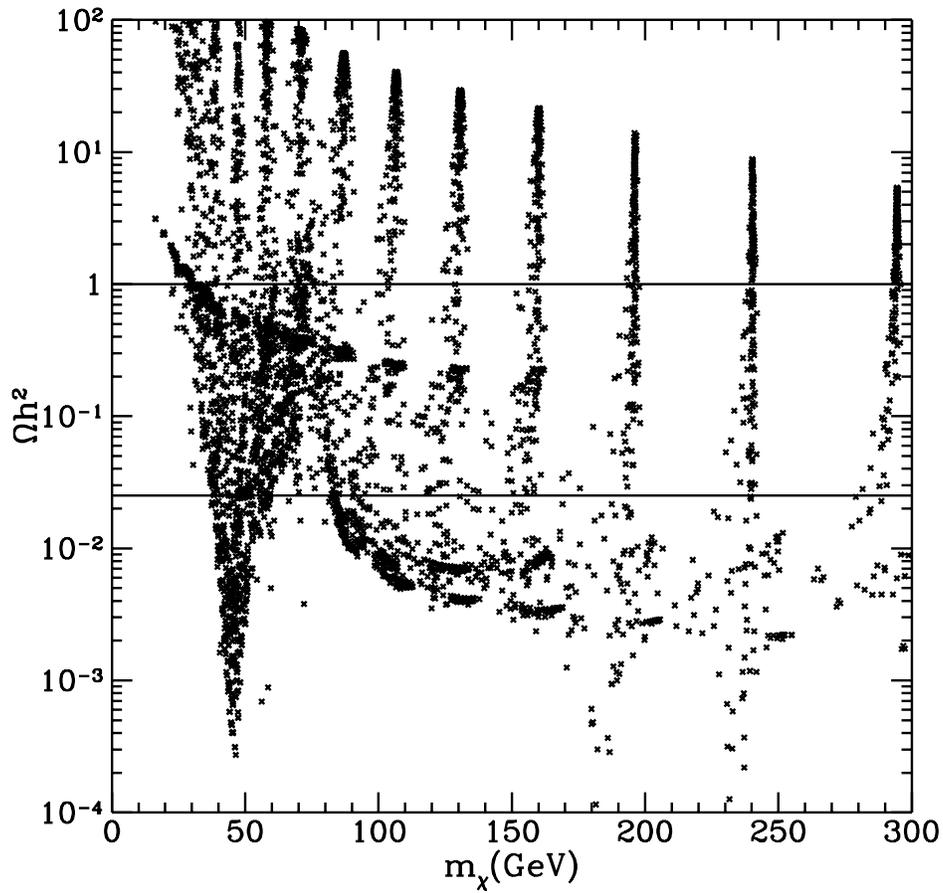,width=5in}}
\caption{Cosmological abundance of a WIMP versus the WIMP mass.
     Each point represents the result for a given choice of the
     MSSM parameters.  The spikes arise simply as a consequence
     of our method of tiling the SUSY parameter space---they
     have no physical significance.  From Ref.~\protect\citenum{jkg}.}
\label{fig:relicabundance}
\end{figure}

This is
illustrated in Fig.~\ref{fig:relicabundance} where the
cosmological abundance $\Omega_\chi$ (times $h^2$) is plotted
versus the neutralino mass $m_\chi$.  Each point represents
one supersymmetric model, or equivalently, one choice of the
MSSM parameters.  Models with $\Omega_\chi h^2 \ga 1$ are
excluded if the Universe is at least 10 Gyr old,
while those with $\Omega_\chi h^2 \la 0.025$ are
cosmologically consistent, but probably give too few neutralinos
to account for the dark matter in galactic halos.  The numerous
models in which the neutralino abundance is between these two
limits provide excellent dark-matter candidates.

\subsection{Direct Detection of WIMPs}

SUSY particles are now the
primary targets of the next generation of accelerator
experiments.  However, one can also try to detect neutralinos in
the Galactic halo.  In order to account for the dynamics of the
Milky Way, the {\it local} dark-matter density must be $\rho_0
\simeq 0.4\, {\rm GeV}/{\rm cm}^3$, and whatever particles or
objects make up the dark-matter halo must be moving with a
velocity dispersion of 270 km/sec.

Perhaps the most promising technique to detect WIMPs is
detection of the ${\cal O}(30\, {\rm keV})$ nuclear recoil
produced by elastic scattering of neutralinos from nuclei in
low-background
detectors.\cite{witten,kim,labdetectors}  A particle with mass
$m_\chi\sim100$ GeV and electroweak-scale interactions
will have a cross section for elastic scattering from a nucleus
which is $\sigma \sim 10^{-38}\,{\rm cm}^2$.  If the local halo
density is $\rho_0\simeq0.4$ GeV~cm$^{-3}$, and the particles
move with velocities $v\sim 300$ km~sec$^{-1}$, then the rate
for elastic scattering of these particles from, e.g., germanium
which has a mass $m_N \sim70$ GeV, will be $R \sim \rho_0
\sigma v / m_\chi/m_N \sim1$ event~kg$^{-1}$~yr$^{-1}$.  If a
$100$-GeV WIMP moving at $v/c\sim10^{-3}$ elastically scatters
with a nucleus of similar mass, it will impart a recoil energy
up to 100 keV to the nucleus.  Therefore, if we have 1 kg of
germanium, we expect to see roughly one nucleus per year
spontaneously recoil with an energy of ${\cal O}(30 {\rm keV})$.

More precise calculations of the detection rate include the
proper neutralino-quark interaction, the QCD and
nuclear physics that turn a neutralino-quark
interaction into a neutralino-nucleus interaction, and a full
integration over the WIMP velocity distribution.  Even if all of these
physical effects are included properly, there is still some
uncertainty in the predicted event rates that arises from
current limitations in our understanding of, e.g., squark,
slepton, chargino, and neutralino masses and mixings.
Therefore, rather than
make a single precise prediction, theorists generally survey the
available SUSY parameter space.  Doing so, one finds event rates between
$10^{-4}$ to 10 events~kg$^{-1}$~day$^{-1}$ (Ref.~\citenum{jkg}), as shown
in Fig.~55 of Ref.~\citenum{jkg}, although there may be models
with rates that are a bit higher or lower.

\subsection{Energetic Neutrinos from WIMP Annihilation}

Energetic neutrinos from WIMP annihilation in the Sun and/or
Earth provide an alternative avenue for indirect detection of
WIMPs.\cite{SOS}
If, upon passing through the Sun, a WIMP scatters elastically from a
nucleus therein to a velocity less than the escape velocity, it
will be gravitationally bound to the Sun.  This leads to a
significant enhancement in the density of WIMPs in the center of
the Sun---or by a similar mechanism, the Earth.  These WIMPs
will annihilate to, e.g., $c$, $b$, and/or $t$ quarks, and/or gauge and
Higgs bosons.  Among the decay products of these particles
will be energetic muon neutrinos which can escape from the
center of the Sun and/or Earth and be detected in neutrino
telescopes such as IMB, Baksan, Kamiokande, MACRO, or AMANDA.
The energies of these muons will be typically 1/3 to 1/2 the
neutralino mass (e.g., 10s to 100s of GeV) so they will be much
more energetic than ordinary solar neutrinos (and therefore
cannot be confused with them).\cite{JunKam95}  The signature of such a
neutrino would be the Cerenkov radiation emitted by an upward
muon produced by a charged-current interaction between the
neutrino and a nucleus in the rock below the detector.

The annihilation rate of these WIMPs equals the rate for
capture of these particles in the Sun, which can
be calculated.\cite{pressspergel}  The flux of
neutrinos at the Earth depends also on the Earth-Sun distance,
WIMP annihilation branching ratios, and the decay branching
ratios of the annihilation products.  The flux of upward muons
depends on the flux of neutrinos and the cross section for
production of muons, which depends on the square of the neutrino
energy.  

As in the case of direct detection, the precise
prediction involves numerous factors from particle and nuclear
physics and astrophysics, and on the SUSY parameters.
When all these factors are taken into account, predictions for
the fluxes of such muons in SUSY models
seem to fall for the most part between $10^{-6}$ and 1
event~m$^{-2}$~yr$^{-1}$ (Ref.~\citenum{jkg}), as shown in
Fig.~57 of Ref.~\citenum{jkg}, although the numbers may be a bit
higher or lower in some models.  Presently, IMB, Kamiokande,
Baksan, and MACRO constrain the flux of energetic neutrinos from
the Sun to be $\lesssim0.02$~m$^{-2}$~yr$^{-1}$
(Ref.~\citenum{kamiokande,imb,baksan,macronew}).  Larger and more
sensitive detectors such as super-Kamiokande\cite{superK} and
AMANDA\cite{AMANDA} are now operating, and others are being
constructed.\cite{IceCube}

\subsection{Recent Results}

\begin{figure}[htbp]
\centerline{\psfig{file=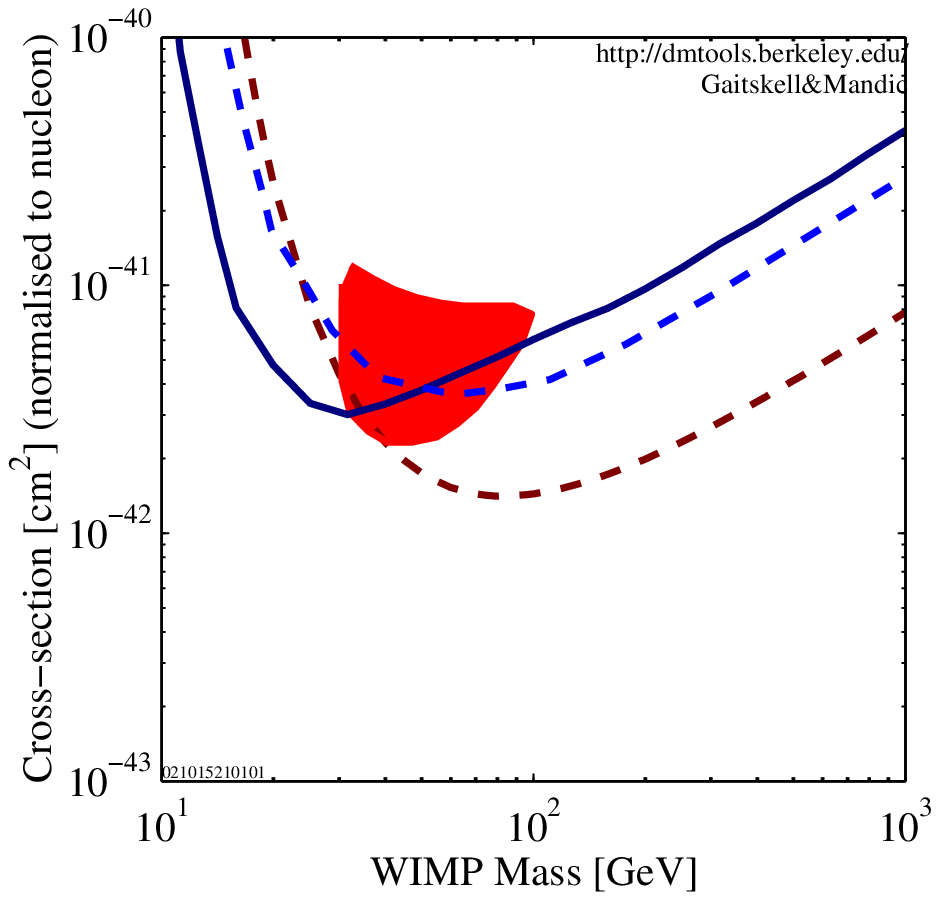,width=5in}}
\centerline{\psfig{file=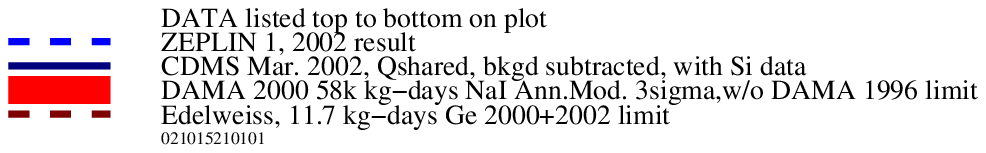,width=5in}}
\caption{The regions of the WIMP mass--cross-section parameter
     space for a WIMP with scalar interactions with nucleons.
     The shaded area is the parameter space inferred from the
     DAMA modulation, and the curves show upper limits from
     CDMS, EDELWEISS, and ZEPLIN.  (From {\tt http://dmtools.berkeley.edu}.)}
\label{fig:DAMA}
\end{figure}

There has been some controversy and excitement among
dark-matter experimentalists in recent years.  The DAMA
collaboration\cite{dama} has for several years seen an annual
modulation in the event rate in their NaI detector, which they
attribute to a WIMP.
A WIMP can interact with nuclei either through a scalar
interaction (where the WIMP-nucleus cross section scales
with the nuclear mass), or through an axial-vector interaction
(where the WIMP-nucleus cross section depends on
something like the nuclear spin or magnetic moment).
If the DAMA modulation is attributed to a WIMP with a scalar
interaction with nuclei, then it implies a WIMP mass and
WIMP-nucleon cross section in the region indicated in
Fig.~\ref{fig:DAMA}.  For several years, null searches in the
CDMS Ge detector\cite{cdms} ruled out most of this region.
This past year, new null results from EDELWEISS\cite{edelweiss}
and ZEPLIN\cite{zeplin} seem to have now ruled out the entire
DAMA parameter space.

\begin{figure}[htbp]
\centerline{\psfig{file=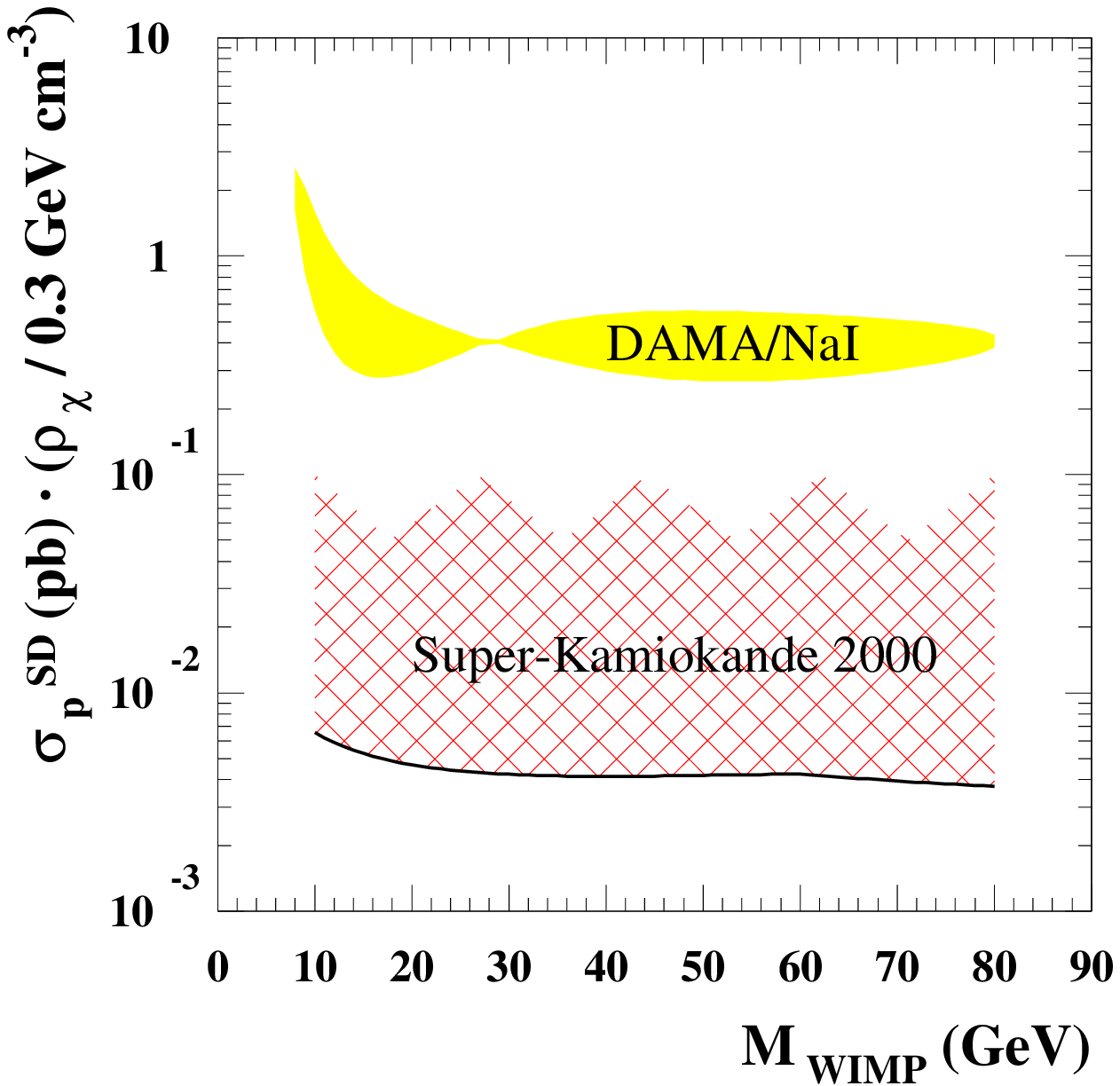,width=5in}}
\caption{The regions of the WIMP mass--cross-section parameter
     space for a WIMP with axial-vector interactions with
     protons.  The shaded region shows the parameter space
     inferred if the DAMA modulation is attributed to a WIMP
     with only a spin-dependent interaction with protons.  The
     lower curve is the upper limit from Super-Kamiokande.  From
     Ref.~\protect\citenum{piero}.}
\label{fig:proton}
\end{figure}

\begin{figure}[htbp]
\centerline{\psfig{file=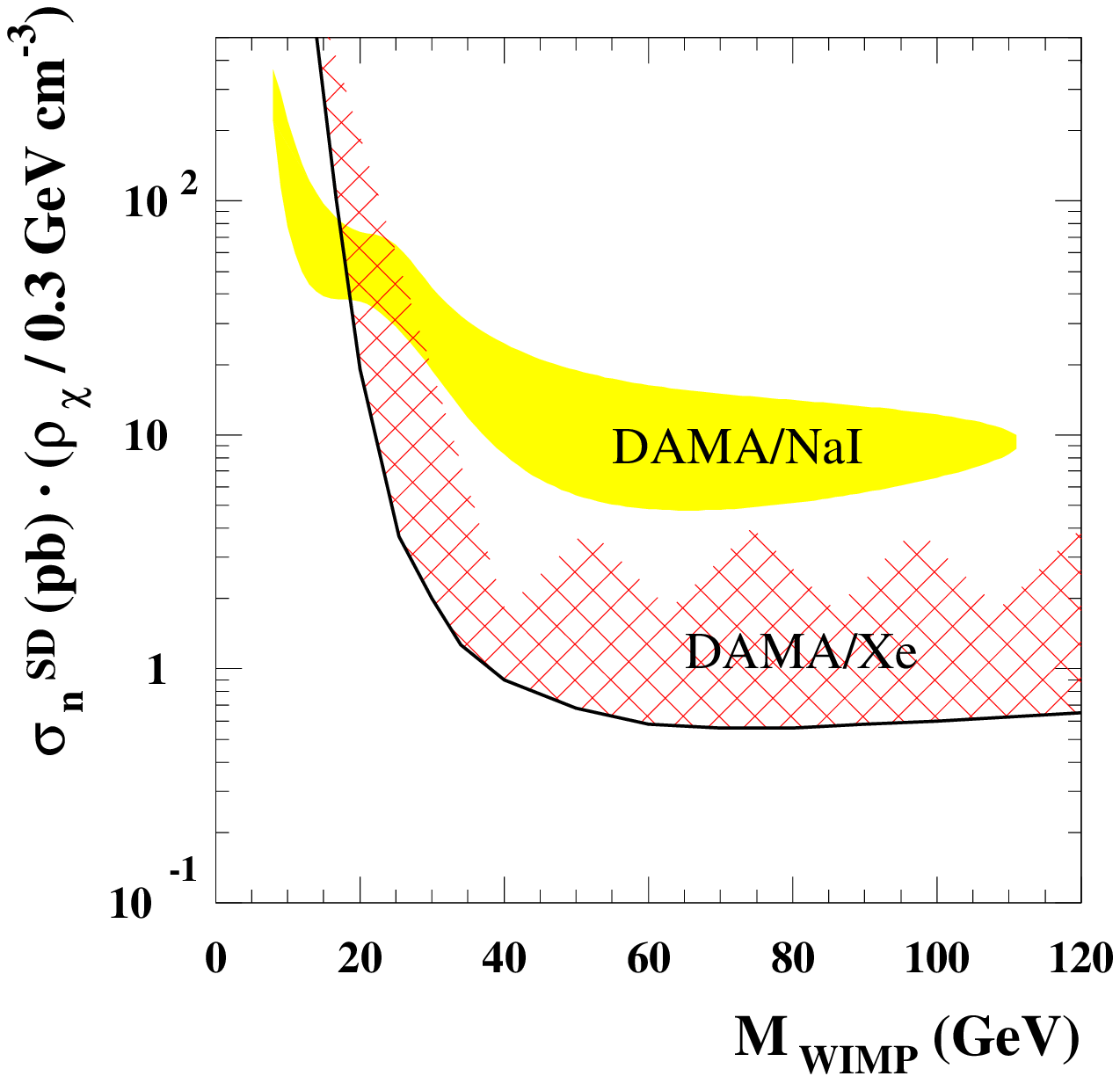,width=5in}}
\caption{The regions of the WIMP mass--cross-section parameter
     space for a WIMP with axial-vector interactions with
     neutrons.  The shaded region shows the parameter space
     inferred if the DAMA modulation is attributed to a WIMP
     with only a spin-dependent interaction with neutrons.  The
     lower curve is the upper limit from the DAMA Xe
     experiment.  From Ref.~\protect\citenum{piero}.}
\label{fig:neutron}
\end{figure}

But what if the WIMP has an axial-vector interaction?  Both Na
and I have a spin carried primarily by an unpaired proton.
Although $^{73}$Ge also has a spin carried primarily by a
proton, its isotopic abundance is only 7\%.  Thus, if the DAMA
modulation is due to a WIMP with a spin-dependent interaction
with protons, it would evade detection in CDMS\cite{piero} (and
probably also in EDELWEISS; a careful analysis 
is now in progress\cite{kurylov}).  In Ref.~\citenum{piero}, however,
we showed using the model-independent analysis of
Refs.~\citenum{modelindependent} that if the DAMA modulation were
attributed to a WIMP-proton axial-vector interaction, then these
WIMPs would accumulate efficiently in the Sun (which is made
primarily of protons) and annihilate therein.  The resulting
flux of energetic neutrinos would have been well over an order
of magnitude larger than the current upper bounds from 
Kamiokande, Baksan, and MACRO, as shown in
Fig.~\ref{fig:proton}.  Alternatively, a small fraction of the
nuclear spins in Na and I could be due to neutrons, and if so,
the DAMA modulation could be explained by a WIMP with a
spin-dependent interaction with neutrons.  In this case,
however, the WIMP-neutron interaction would have to be quite
strong, and the WIMP would have already shown up in another of
DAMA's detectors that is made of Xe, as shown in
Fig.~\ref{fig:neutron}, and probably also in ZEPLIN (a
detailed analysis is now in progress\cite{kurylov}).

\begin{figure}[htbp]
\centerline{\psfig{file=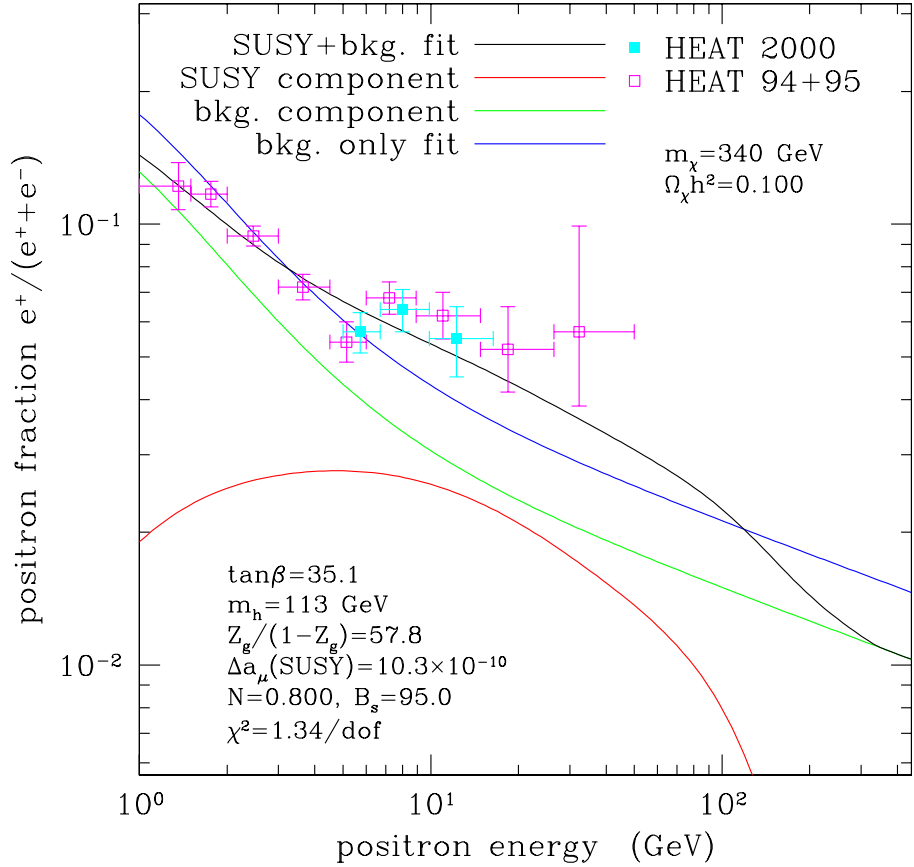,width=5in}}
\caption{The differential positron flux divided by the sum of
     the differential electron-plus-positron flux.  Shown are
     the HEAT data as well as theoretical models of background
     and the background with a supersymmetric dark-matter
     annihilation signal added.  From Ref.~\protect\citenum{baltz}.}
\label{fig:positrons}
\end{figure}

\subsection{WIMPs and Exotic Cosmic Rays}

WIMPs might also be detected via observation of exotic
cosmic-ray positrons, antiprotons, and gamma rays produced by
WIMP annihilation in the Galactic halo.  The difficulty
with these techniques is discrimination between WIMP-induced
cosmic rays and those from traditional astrophysical
(``background'') sources. However, WIMPs may
produce distinctive cosmic-ray signatures.  As illustrated in
Fig.~\ref{fig:positrons},\cite{baltz} WIMP annihilation might produce a
cosmic-ray-positron excess at high energies.\cite{baltz,positrons}
There are now several balloon (e.g., BESS,
CAPRICE, HEAT, IMAX, MASS, TS93) and satellite (AMS and PAMELA)
experiments that have recently flown or are about to be flown to search for
cosmic-ray antimatter.  In fact, the HEAT experiment may already
show some evidence for a positron excess at high
energies.\cite{heatpositrons}

WIMP annihilation will produce an antiproton excess at low
energies,\cite{JunKam94} although Ref.~\citenum{pieropbar} claims
that traditional astrophysical sources can mimic such an excess.
They argue that the antiproton background 
at higher energies ($\ga$few GeV) is better understood, and that
a search for an excess of these higher-energy antiprotons would
thus provide a better WIMP signature.  

Direct WIMP annihilation to two photons can produce a gamma-ray
line, which could not be mimicked by a traditional astrophysical
source, at an energy equal to the WIMP mass.  WIMPs could also
annihilate directly to a photon and a $Z^0$
boson,\cite{berkap,pieroZphoton} and these photons will be
monoenergetic with an energy that differs from that of the
photons from direct annihilation to two  photons.  Resolution of
both lines and measurement of their
relative strengths would shed light on the composition of the
WIMP.  Ground-based experiments like STACEE or CELESTE or the
GLAST satellite will seek this annihilation radiation.

It was recently argued\cite{GS} that there may be a very
dense dark-matter spike, with a dark-matter density that scales
with radius $r$ as $\rho(r) \propto r^{-2.25}$ from the Galactic
center, around the black hole at the Galactic center.  If so, it
would give rise  to a huge flux of annihilation radiation.
However, others have questioned whether this spike really arises.\cite{zhao}

\subsection{Axions}

The other leading dark-matter candidate is the
axion.\cite{axion}  The QCD Lagrangian may be written 
\begin{equation}
     {\cal L}_{QCD} = {\cal L}_{\rm pert} + \theta {g^2 \over 32
     \pi^2} G \widetilde{G},
\end{equation}
where the first term is the perturbative Lagrangian responsible
for the numerous phenomenological successes of QCD.  However,
the second term (where $G$ is the gluon field-strength tensor
and $\widetilde{G}$ is its dual), which is a consequence of
nonperturbative effects, violates $CP$.  From constraints to the
neutron electric-dipole moment, $d_n \la 10^{-25}$ e~cm, it can
be inferred that $\theta \la 10^{-10}$.  But why is $\theta$ so
small?  This is the strong-$CP$ problem.

The axion arises in the Peccei-Quinn (PQ) solution to the strong-$CP$
problem,\cite{PQ} which twenty-five years after it was
proposed still seems to be the most promising solution.  
A global $U(1)_{PQ}$ symmetry broken at a
scale $f_{PQ}$, and $\theta$ yields a dynamical field which is
the Nambu-Goldstone mode of this symmetry.  At temperatures
below the QCD phase transition, nonperturbative quantum effects
break explicitly the symmetry and drive $\theta\rightarrow 0$.
The axion is the pseudo-Nambu-Goldstone boson of this
near-global symmetry.  Its mass is $m_a \simeq\, {\rm
eV}\,(10^7\, {\rm GeV}/ f_a)$, and its coupling to ordinary
matter is $\propto f_a^{-1}$.

The Peccei-Quinn solution works equally well for
any value of $f_a$.  However, a variety
of astrophysical observations and laboratory experiments
constrain the axion mass to be $m_a\sim10^{-4}$ eV.
Smaller masses would lead to an
unacceptably large cosmological abundance.  Larger masses
are ruled out by a combination of constraints from supernova
1987A, stellar evolution, laboratory experiments, and a search
for two-photon decays of relic axions.

Curiously enough, if the axion mass is in the relatively small viable
range, the relic density is $\Omega_a\sim1$ and may therefore
account for the halo dark matter.  Such axions would be produced
with zero momentum by a misalignment mechanism in the early
Universe and therefore act as cold dark matter.  During the process of
galaxy formation, these axions would fall into the Galactic
potential well and would therefore be present in our halo with a
velocity dispersion near 270 km~sec$^{-1}$.

If $m_a\sim10^{-4}$ eV, the magnitude of the explicit symmetry
breaking is incredibly tiny compared with the PQ scale, so the
global PQ symmetry, although broken, must be very close to exact.
There are physical arguments involving, for example, the
nonconservation of global charge in evaporation of a black hole
produced by collapse of an initial state with nonzero global
charge, that suggest that  global symmetries should be violated
to some extent in quantum gravity.  In order for the PQ
mechanism to work for $m_a\sim10^{-4}$, the coupling of a
generic global-symmetry-violating term from quantum-gravity effects
must be extraordinarily small (e.g., $\la
10^{-55}$).\cite{gravity}  Of course, we have at this point no
predictive
theory of quantum gravity, and several mechanisms for forbidding
these global-symmetry violating terms have been
proposed.\cite{solutions}  Therefore, these arguments by no
means ``rule out'' the axion solution.  Rather, discovery of an
axion would provide much needed clues to the nature of
Planck-scale physics.

\begin{figure}[htbp]
\centerline{\psfig{file=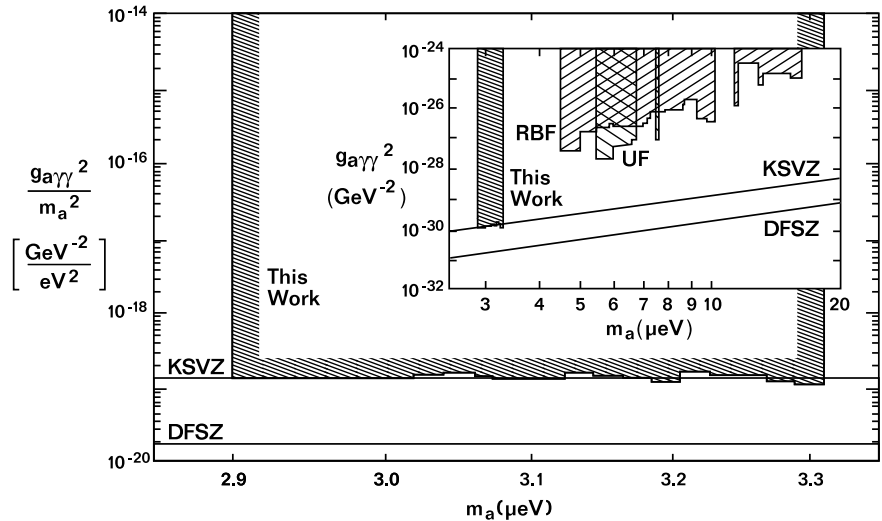,width=5in}}
\caption{Regions of axion mass-coupling parameter space
     currently being probed by an ongoing search at
     Livermore\protect\cite{axionexperiments}.}
\label{fig:axion}
\end{figure}

There is a very weak coupling of an axion to photons through the
anomaly.  The axion can therefore decay to two
photons, but the lifetime is $\tau_{a\rightarrow \gamma\gamma}
\sim 10^{50}\, {\rm s}\, (m_a / 10^{-5}\, {\rm eV})^{-5}$ which
is huge compared to the lifetime of the Universe and therefore
unobservable.  However, the $a\gamma\gamma$ term in the
Lagrangian is ${\cal L}_{a\gamma\gamma} \propto a {\vec E} \cdot
{\vec B}$ where ${\vec E}$ and ${\vec B}$ are the electric and
magnetic field strengths.  Therefore, if one immerses a resonant
cavity in a strong magnetic field, Galactic axions that pass
through the detector may be converted to fundamental excitations
of the cavity, and these may be observable.\cite{sikivie}  Such
an experiment is currently underway\cite{axionexperiments} and
has already begun to probe part of the cosmologically
interesting parameter space (see Fig. \ref{fig:axion}), and it
should cover most of the interesting region parameter space in
the next few years. A
related experiment, which looks for excitations of Rydberg
atoms, is also seeking dark-matter axions.\cite{rydberg}
Although the sensitivity of this technique should be
excellent, it can only cover a limited axion-mass range.  The
CERN Axion Solar Telescope (CAST) project\cite{CAST} is
searching for $m_a\simeq O({\rm eV})$ axions by looking for
resonant conversion of thermal axions from the Sun into x rays.
This mass range is available only if there are loopholes in the
stellar-evolution calculations that nominally exclude these
masses.

\subsection{Self-Interacting Dark Matter?}

N-body simulations of structure formation with collisionless
dark matter show dark-matter cusps, density profiles that fall
as $\rho(r)\propto 1/r$ with radius $r$ near the galactic
center,\cite{nfw} while some dwarf-galaxy rotation curves
indicate the existence of a density core in their
centers.\cite{moore}  This has prompted some theorists to
consider
self-interacting dark matter.\cite{ss}  If dark-matter
particles elastically scatter from each other in a galactic
halo, then heat can be transported from the halo center to the
outskirts; in this way, the cusp can be smoothed into a core.
In order for this mechanism to work, however, the
elastic-scattering cross section must be $\sigma_{\rm el} \sim
10^{-(24-25)} (m_\chi/{\rm GeV})$ cm$^2$, roughly thirteen
orders of magnitude larger than the cross section expected for
WIMPs, and even further from that for axions.  If the cross
section is stronger, the halo will undergo core
collapse,\cite{corecollapse} and if it is weaker, the heat
transport is not sufficiently efficient to remove the
dwarf-galaxy dark-matter cusp.

The huge discrepancy between the magnitude of the required
scattering cross section and that for WIMPs and axions has made
self-interacting dark matter unappealing to most WIMP and axion
theorists (but see, e.g., Refs.~\citenum{mcdonald}).  However,
theoretical prejudices aside, self-interacting dark matter now
seems untenable observationally.  If dark matter is collisional,
dark-matter cores should equilibrate and become round.
Non-radial arcs in the gravitational-lensing system MS2137-23
require a non-spherical core and thus rule out the scattering
cross sections required to produce dwarf-galaxy
cores.\cite{jordi}  One possible loophole is that the
scattering cross section is inversely proportional to the
relative velocity of the scattering particles; this would
lengthen the equilibration time in the core of the cluster
MS2137-23.  This possibility has now been ruled out, however, by
x-ray observations of the giant elliptical galaxy NGC 4636 which
shows a very dense dark-matter cusp at very small
radii.\cite{loewenstein}

\section{The Cosmic Microwave Background, Large-Scale Structure,
and Inflation}

\subsection{Recent Progress in the CMB}

In the past few years, the cosmic microwave background (CMB) has
begun to provide perhaps the most exciting opportunity for learning
about new physics at ultra-high-energy scales (for recent
reviews, see, e.g., Refs.~\citenum{KamKos99,HuDod01,church}).
We have already seen spectacular advances in
measurements of temperature fluctuations in the
CMB\cite{Miletal99,deBetal00,Hanetal00,dasiT,Masetal02}
that have led to major advances in our ability to
characterize the largest-scale structure of the Universe, the
origin of density perturbations, and the early Universe.  In
just the past few months we have seen the first detection of CMB
polarization\cite{dasi} and a spectacular measurement of
fluctuations from 10-degree to sub-degree angular
scales.\cite{archeops}  In the next few months we
should see even more improvements from the MAP
satellite,\cite{map} and even more with the launch of the Planck
satellite\cite{PLANCK} in 2007.

The primary aim of these experiments has been to determine the CMB
power spectrum, $C_\ell$, as a function of multipole moment $\ell$.
Structure-formation theories predict a series of bumps in the power
spectrum in the region $50\lesssim \ell \lesssim 1000$, arising
from oscillations in the baryon-photon fluid before CMB
photons last scatter.\cite{peaks}  The rich structure in
these peaks allows simultaneous determination of the geometry of the
Universe,\cite{KamSpeSug94,Junetal96a} the baryon density,
Hubble constant,
matter density, and cosmological constant, as well as the nature
(e.g., adiabatic, isocurvature, or topological defects) and spectrum
of primordial perturbations.\cite{Junetal96b}

\begin{figure}[htbp]
\centerline{\psfig{file=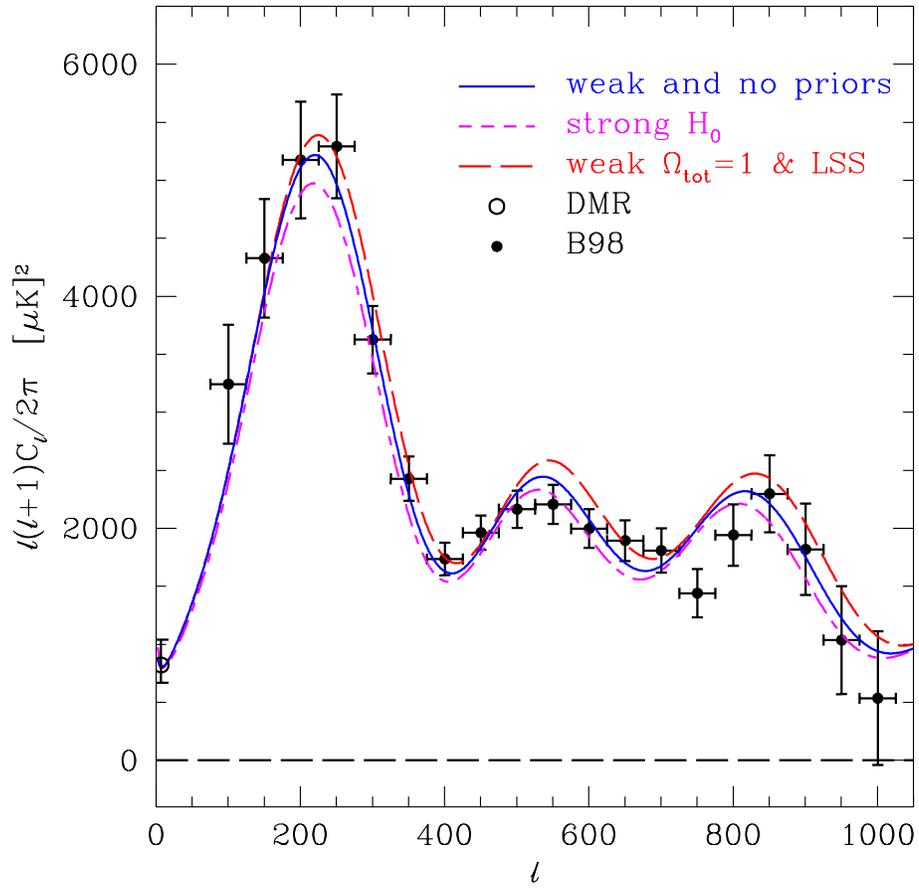,width=5in}}
\caption{The CMB power spectrum measured
     recently by BOOMERanG\protect\cite{boom}.  Similar results have been
     obtained also by DASI\protect\cite{dasiT} and
     MAXIMA.\protect\cite{max}}
\label{fig:boom}
\end{figure}

\begin{figure}[htbp]
\centerline{\psfig{file=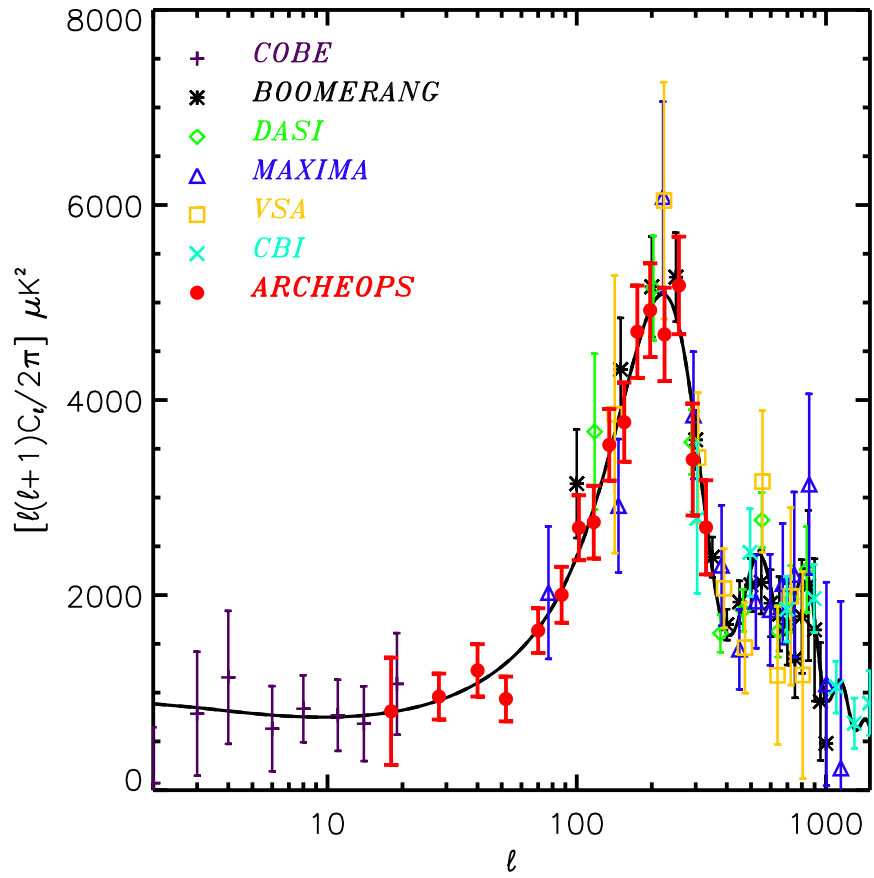,width=5in}}
\caption{The CMB power spectrum measured
     recently by ARCHEOPs, as well as by previous experiments.
     From Ref.~\protect\citenum{archeops}.}
\label{fig:archeops}
\end{figure}

Within the past two years, three independent experiments that use
different techniques, observing strategies, and frequencies have each
measured the power spectrum in the range
$50\lesssim\ell\lesssim1000$ (Refs.~\citenum{boom,max,dasi}), and the
existence of the second and third peaks has now been
confirmed,\cite{boom} as shown in Fig.~\ref{fig:boom}.  These
experiments represent a watershed event in cosmology, as they suggest for
the first time that the Universe is flat and that structure grew from
a nearly scale-invariant spectrum of primordial density perturbations.
These two properties are robust predictions of
inflation,\cite{inflation,perturbations} a period of accelerated expansion in
the very early Universe driven by the vacuum energy associated
with some new, yet undetermined, ultra-high-energy physics.  The
new results from ARCHEOPS,\cite{archeops} shown in
Fig.~\ref{fig:archeops}, overlap and also interpolate between
the largest-angle measurements from COBE and the degree-scale
experiments.  The agreement with the earlier experiments is
stunning.

\begin{figure}[htbp]
\centerline{\psfig{file=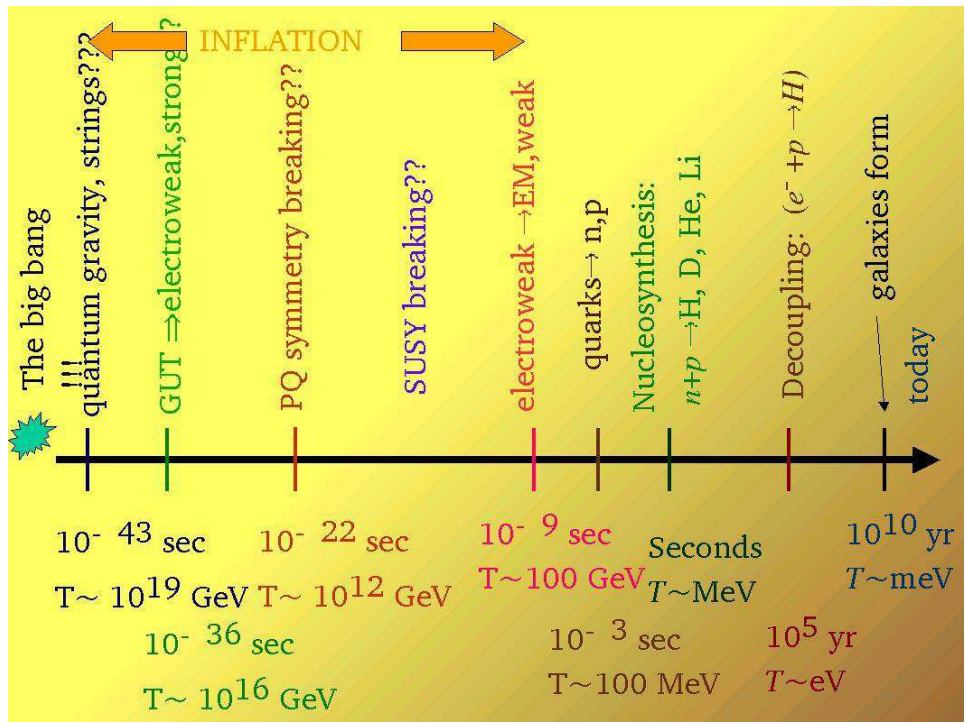,width=5in}}
\caption{Logarithmic history of the
     Universe.}
\label{fig:timeline}
\end{figure}

Although these recent CMB tests suggest that we are on the right 
track with inflation, we still have no idea what new physics may 
have given rise to inflation.  Plausible theoretical models
place the energy scale of inflation anywhere from the
Planck scale to the electroweak scale, and associate the
inflaton (the scalar field responsible for inflation) with new
fields that arise in string theory, GUTs, the Peccei-Quinn
mechanism, supersymmetry breaking, and electroweak-scale
physics, as shown in Fig.~\ref{fig:timeline}.

\subsection{Inflation, Gravitational Waves, and CMB Polarization}

Perhaps the most promising avenue toward further tests of
inflation as well as determination of the energy scale of
inflation is the gravitational-wave background.
Inflation predicts that quantum fluctuations in the
spacetime metric during inflation should give rise to a
stochastic gravitational-wave background with a
nearly-scale-invariant spectrum (defined to be the fourth root
of the inflaton potential during inflation).\cite{AbbWis84}
Inflation moreover predicts
that the amplitude of this gravitational-wave background should
be proportional to the square of the energy scale of inflation.

These gravitational waves will produce temperature fluctuations
at large angles.  Upper limits to the amplitude of large-angle
temperature fluctuations already constrain the energy scale of
inflation to be less than $3\times10^{16}$ GeV.  However, since
density perturbations can also produce such temperature
fluctuations, observed temperature fluctuations cannot alone be
used to detect the gravitational-wave background.

Instead, progress can be made with the polarization of the CMB.
Both gravitational waves and density perturbations will produce
linear polarization in the CMB, and the two polarization patterns
differ.  More precisely, gravitational waves
produce polarization with a distinctive curl pattern
that cannot be mimicked by density perturbations (at linear
order in perturbation theory; see
below).\cite{KamKosSte97,SelZal97}  Moreover, inflation
robustly predicts that the amplitude of this curl
depends on the square of the energy scale of inflation.

Is this signal at all detectable?  If the energy scale 
of inflation is much below the GUT scale, then the polarization
signal will likely be too small to ever be detected.  However,
if inflation had something to do with GUTs---as many, if not
most theorists believe---then the signal is conceivably
detectable by a next-generation CMB experiment.\cite{KamKos98}
Although the MAP satellite, launched just last month, is unlikely to
have sufficient sensitivity to detect the curl component from
inflationary gravitational waves, the Planck satellite, a
European Space Agency experiment to be launched in 2007, 
should have sufficient sensitivity to detect the CMB curl
component as long as the energy scale of inflation is greater
than roughly $5\times10^{15}$ GeV.  However, Planck will not be
the end of the line.  An experiment that integrates more deeply
on a smaller region of sky can improve the sensitivity to the
inflationary gravitational-wave background by almost two orders of
magnitude.\cite{JafKamWan}  Moreover, there are several very promising ideas
being pursued now that could improve the detector sensitivity by 
more than an order of magnitude within the next decade.  Putting 
these two factors together, it becomes likely that a CMB
polarization experiment that probes inflationary energy
scales to below $10^{15}$ GeV---and thus accesses the entire
favored GUT parameter space---could be mounted on a ten-year
timescale (if not sooner).

\subsection{Cosmic Shear and the CMB}

There is, however, another source of a curl component.  Cosmic
shear (CS)---weak gravitational lensing of the CMB due to
large-scale structure along the line of sight---results in a
fractional conversion of the gradient mode from density
perturbations to the curl component.\cite{ZalSel98}
The CS-induced
curl thus introduces a noise from which IGWs must be distinguished.
If the IGW amplitude (or $E_{\rm infl}$) is sufficiently large,
the CS-induced curl will be no problem.  However, as $E_{\rm
infl}$ is reduced, the IGW signal becomes smaller and will at
some point get lost in the CS-induced noise.  If it is not
corrected for, this confusion leads to a minimum detectable IGW
amplitude.\cite{LewChaTur02,KesCooKam02,KnoSon02}

\begin{figure}[htbp]
\centerline{\psfig{file=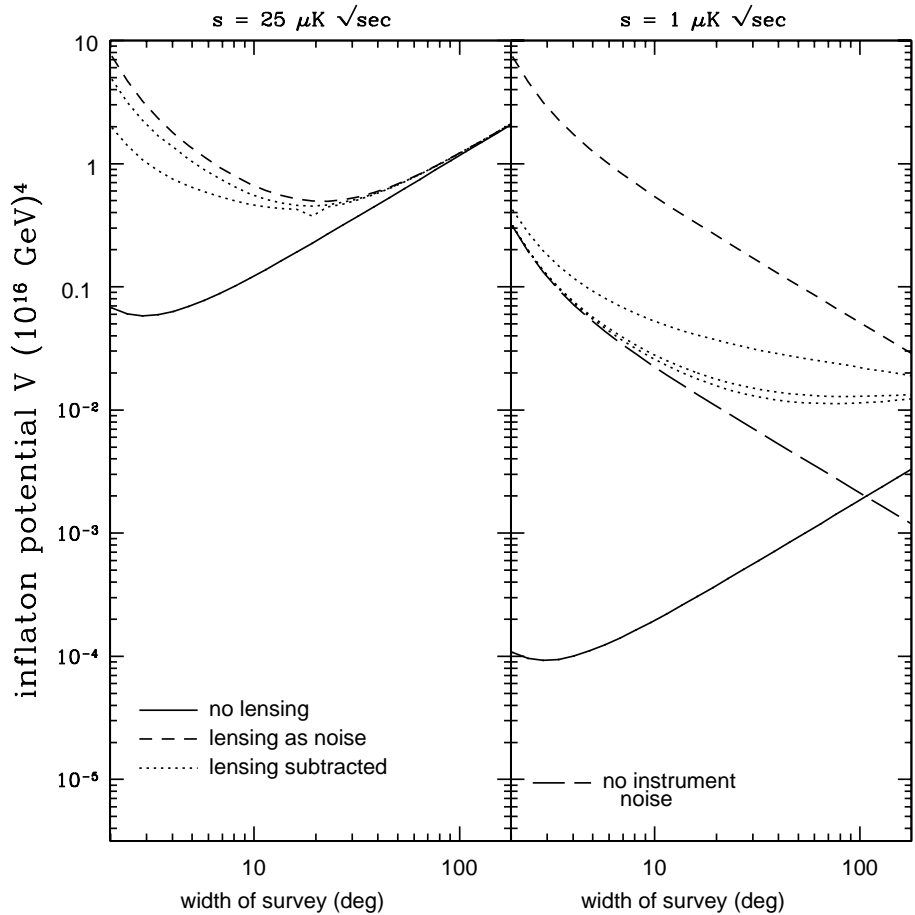,width=5in}}
\caption{Minimum inflation potential observable at
     $1\sigma$ as a function of survey width for a one-year
     experiment.  The left panel shows an experiment with
     noise-equivalent temperature (NET, the detector sensitivity)
     $s=25\, \mu{\rm K}~\sqrt{\rm sec}$.
     The solid curve shows results that could be obtained if
     there were no cosmic shear.  The long-dash curve shows the
     minimum detectable inflaton-potential height with the
     optimal detection strategy.  See
     Ref.~\protect\citenum{KesCooKam02} for more details and a full
     description of the other curves.}
\label{fig:kesden}
\end{figure}

In addition to producing a curl component, CS also introduces
distinct higher-order correlations in the CMB temperature
pattern.  Roughly speaking, lensing can stretch the image of the
CMB on a small patch of sky and thus lead to something akin to
anisotropic correlations on that patch of sky, even though the
CMB pattern at the surface of last scatter had isotropic
correlations.  By mapping these effects, the CS can be
mapped as a function of position on the sky.\cite{SelZal99}
The observed CMB polarization can then be corrected for these
lensing deflections to reconstruct the intrinsic CMB
polarization at the surface of last scatter (in which the only
curl component would be that due to IGWs).

Refs.~\citenum{KesCooKam02,KnoSon02} show that
if the gravitational-wave background is large enough to be
accessible with the Planck satellite, then the cosmic-shear
contribution to the curl component will not get in the way.
However, to go beyond Planck, the cosmic-shear distortion to the
CMB curl will need to be subtracted by mapping the cosmic-shear
deflection with higher-order temperature-polarization
correlations.  Ultimately, if the energy scale is $E_{\rm
infl}\lesssim 2\times10^{15}$ GeV, then there will be an
irreducible cosmic-shear-induced curl, even with higher-order
correlations.  Thus, if the energy scale of inflation is below
this value, the gravitational-wave background will not be
detectable with the CMB polarization.  Either way, the
cosmic-shear distortions to the CMB will be of interest in their
own right, as they probe the distribution of dark matter
throughout the Universe as well as the growth of density
perturbations at early times.  These goals will be important for
determining the matter power spectrum and thus for testing
inflation and constraining the inflaton potential.

\subsection{CMB and Primordial Gaussianity}

Another prediction of inflation is that the distribution of mass 
in the primordial Universe should be a realization of a Gaussian 
random process.  This means that the distribution of temperature 
perturbations in the CMB should be Gaussian and it moreover
implies a precise relation between all of the higher-order
temperature correlation functions and the two-point correlation
function.  These relations can be tested with future precise CMB
temperature and polarization maps.  See Ref.~\citenum{cozumel} for
a brief review.

\subsection{Structure Formation and Inflation}

Large-scale galaxy surveys have 
become a reality, particularly with the advent of the Two-Degree
Field\cite{2dF} and Sloan Digital Sky Surveys.\cite{SDSS}  We
are now mapping the distribution
of galaxies over huge volumes in the Universe.  Moreover, just
over two years ago, four independent groups reported detection of
cosmic shear through the observation of ellipticity correlations in
distant galaxies.\cite{KaiWilLup00}  In the future,
cosmic-shear measurements will map the distribution of matter
(rather than just the luminous matter probed by galaxy surveys)
over large volumes of space .

If the big bang is a cosmic accelerator, subtle
correlations in the debris from the explosion can provide
valuable information on inflation, just as subtle correlations
in jets in accelerator experiments can provide information about 
the collisions that give rise to them.
The primary aims of galaxy surveys and cosmic-shear maps are
determination of the power spectrum $P(k)$ of the cosmological
matter distribution as a function of wavenumber (inverse
distance) $k$.
These measurements are important for the study of inflation, as
inflation relates the amplitude and shape of the power spectrum
$P(k)$ to the inflaton potential $V(\phi)$ as a function of the
value $\phi$ of the inflaton.  Measurements of
$P(k)$ with the CMB at the very largest scales, to intermediate
scales with galaxy surveys, to the smallest scales with
subgalactic structure\cite{KamLid} are now being pursued.
Moreover, as discussed above, inflation predicts very precise relations
between all of the higher-order correlation functions for the
primordial mass distribution and its two-point correlation
function, and these relations can also be tested with the
observed distribution of mass in the Universe today.
The growth of density perturbations via gravitational infall
alters the precise structure of the correlation hierarchy
from the primordial one.  However, it does so in a calculable
way so that the primordial distribution of density perturbations 
(Gaussian as predicted by inflation? or otherwise?) can be
determined from the distribution observed in the Universe
today.\cite{threept}

Information about the primordial distribution of matter can also 
be obtained by studying the abundances and properties of the
rarest objects in the Universe: clusters of galaxies today and
galaxies at high redshift (see, e.g., Ref.~\citenum{verde}).  Such
objects form at rare
($\gtrsim3\sigma$) high-density peaks in the primordial density
field.  Inflation predicts that the distribution of such peaks
should be Gaussian.  If the distribution is non-Gaussian---for
example, skew-positive with an excess of high-density
peaks---then the abundance of these objects can be considerably
larger.  In such skew-positive models, such objects would also
form over a much wider range of redshifts and thus exhibit a
broader range of properties (e.g., sizes, ages, luminosities,
temperatures).\cite{verdetwo}

\section{Dark Energy}

In addition to confirming the predictions of big-bang
nucleosynthesis and the existence of dark matter, the
measurement of classical cosmological parameters has resulted in
a startling discovery over the
past few years: roughly 70\% of the energy density
of the Universe is in the form of some mysterious
negative-pressure ``dark energy''.\cite{Carroll:2000fy}  Supernova
evidence for an accelerating Universe\cite{Peretal99,Rieetal98}
has now been dramatically bolstered by the discrepancy between
the total cosmological density $\Omega_{\rm tot}\simeq1$
indicated by the CMB and dynamical measurements of the
nonrelativistic-matter density $\Omega_m\simeq0.3$.

As momentous as these results are for
cosmology, they may be even more remarkable from the vantage point of
particle physics, as they indicate the existence of new physics
beyond the standard model plus general relativity.  Either
gravity behaves very peculiarly on the very largest scales, and/or
there is some form of negative-pressure dark energy that
contributes 70\% of the energy density of the Universe.
For this dark energy to accelerate the expansion, its
equation-of-state parameter ${w}\equiv p/\rho$ must satisfy
${w}<-1/3$, where $p$ and $\rho$ are the dark-energy pressure
and energy density, respectively.  The simplest guess for this dark
energy is the spatially uniform, time-independent cosmological
constant, for which ${w}=-1$. Another possibility is
quintessence\cite{Caletal98} or
spintessence,\cite{BoyCalKam01} a cosmic scalar field that is
displaced from the minimum of its potential.  Negative pressure is
achieved when the kinetic energy of the rolling field is less than the
potential energy, so that $-1 \le {w} < -1/3$ is possible.  (In
fact, equations of state $w<-1$, which violate the
dominant-energy condition in general relativity, have now been
considered as well.\cite{caldwell})

Although it is the simplest possibility, a cosmological constant
with this value is strange, as quantum gravity would
predict its value to be $10^{120}$ times the observed value, or
perhaps zero in the presence of some symmetry.  
One of the appealing features of dynamical models for dark
energy is that they may be compatible with a true vacuum
energy which is precisely zero, to which the Universe will
ultimately evolve.  

The dark energy was a complete surprise and remains a
complete mystery to theorists, a stumbling block that, if
confirmed, must be
understood before a consistent unified theory can be
formulated.  This dark energy may be a direct remnant of string
theory, and if so, it provides an exciting new window to
physics at the Planck scale.

The obvious first step to understand the nature of this dark
energy is to determine whether it is a true cosmological
constant, or whether its density evolves with time.  This can be 
answered by determining the expansion rate of the Universe as a
function of redshift.  In principle this can be accomplished
with a variety of cosmological observations (e.g.,
quasar-lensing statistics, cluster abundances and properties,
the Lyman-alpha forest, galaxy and cosmic-shear surveys, etc.).
However, the current best bet for determining the expansion
history is with supernova searches, particularly those that can
reach to redshifts $z\gtrsim1$.  
Here, better systematic-error reduction, better theoretical
understanding of supernovae and
evolution effects, and greater statistics, are all required.
Both ground-based (e.g., the
DMT\cite{DMT} or WFHRI\cite{WFHRI}) and space-based (e.g.,
SNAP\cite{SNAP})
supernova searches can be used to determine the expansion
history.  However, for redshifts $z\gtrsim1$, the principal optical
supernova emission (including the characteristic silicon
absorption feature) gets shifted to the infrared which is
obscured by the atmosphere.  Thus, a space-based observatory
appears to be advisable to reliably measure the expansion
history in the crucial high-redshift regime.

\begin{figure}[htbp]
\centerline{\psfig{file=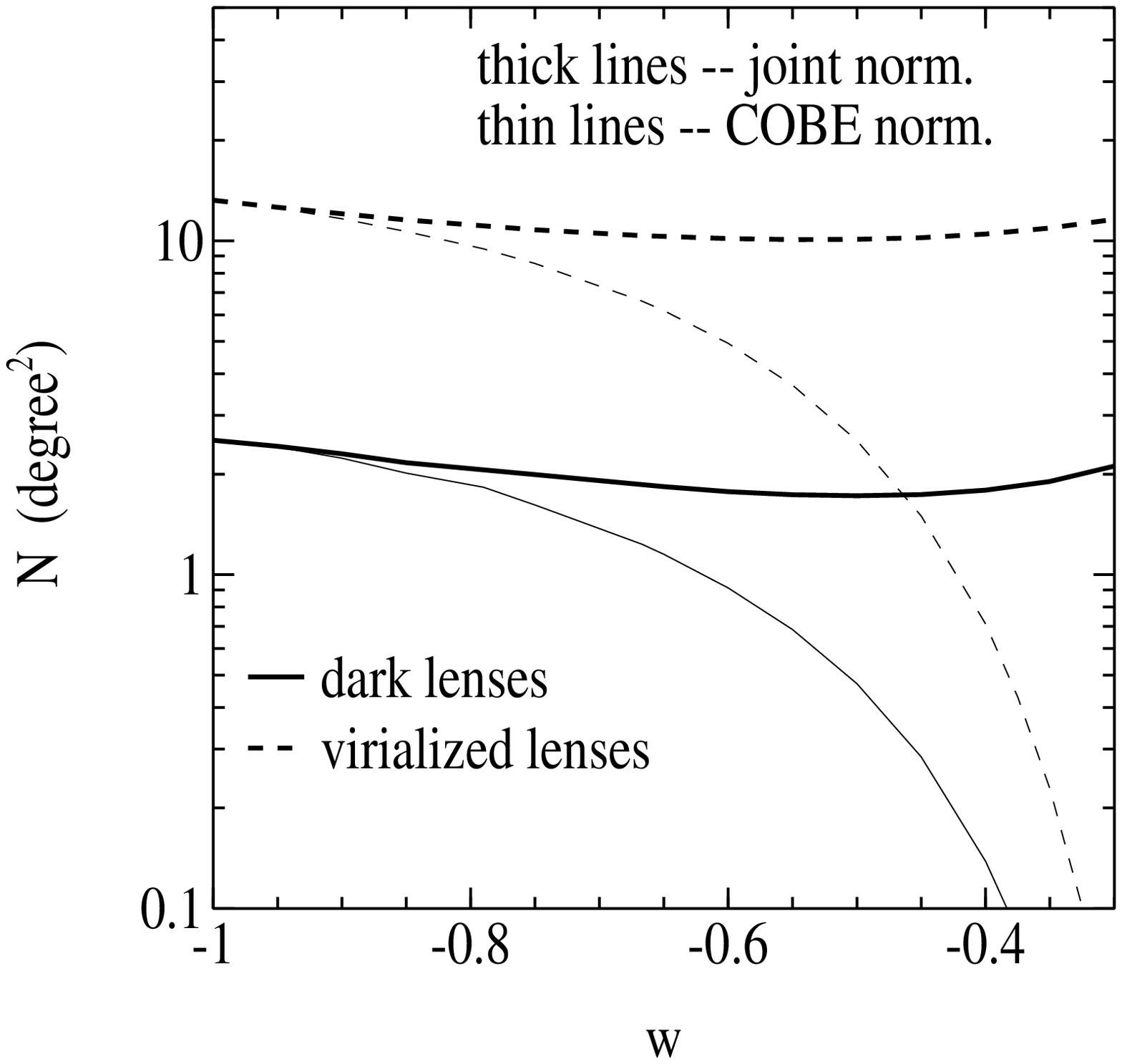,width=4in}}
\caption{The number per square degree of virialized (x-ray
     luminous) and dark(x-ray underluminous) clusters that
     should be detected in in a cosmic-shear survey.  The solid
     and dashed curves are for power spectra normalized to COBE
     and to the cluster abundance.  For each value of $w$, there
     is a unique normalization that fits both COBE and the
     cluster abundance.  See Ref.~\protect\citenum{nevintwo} for more
     details.}
\label{fig:nevin}
\end{figure}

Although supernovae provide perhaps the most direct probe of the
expansion history, there are a number of other indirect probes
as well.   Rather than review them all, I simply discuss, as an
example, one proposal made recently by N. Weinberg and
me\cite{nevinone,nevintwo}.  Wide-angle cosmic-shear surveys of
blank regions
of the sky have already begun, and much larger surveys will soon
to be undertaken.  Individual galaxy clusters should be
detectable in these cosmic-shear maps, but it is possible
that proto-clusters, massive overdensities that are still in the
process of undergoing gravitational collapse, will also appear
in these surveys.  Unlike virialized clusters (i.e., those that
have undergone gravitational collapse), which emit copious
amounts of x-ray radiation, these nonvirialized clusters should
be x-ray underluminous, or appear dark in x-ray bands.  We
showed that the abundance of both virialized and dark clusters
that will be detected in a given cosmic-shear survey, as well as
the ratio of the two, will depend on the equation-of-state
parameter $w$, as shown in Fig. \ref{fig:nevin}.  At least 50
square degrees will need to be surveyed in order for this test
to be carried out.

Most discussions of observational probes of dark energy have
involved the effects of quintessence on the expansion history.
However, if a quintessence field exists, it may have some
couplings to ordinary matter.  If so, it could give rise to
other observable.consequences.  In particular, if the cosmological
``constant'' evolves with time (i.e., is quintessence), then there is 
a preferred frame in the Universe.  Couplings of elementary
particles to the quintessence field may thus give rise to small
apparent violations of Lorentz and/or CPT symmetry (see, e.g.,
Ref.~\citenum{Carroll}).  A variety of accelerator and astrophysical
experiments\cite{Carroll,LueWanKam98} can be done to search for
such exotic signatures.

\section{Summary and Conclusions}

Particle astrophysics and cosmology now represent a very broad
and active research front in non-accelerator probes of new
physics beyond the standard model.  Here I have reviewed
dark-matter searches, the relation between observations of the
cosmic microwave background, the current cosmological mass
distribution, and the early Universe, and the dark-energy
problem.  There are a number of related topics that I did not
discuss, such as neutrino astrophysics, the cosmology of extra
large dimensions, cosmic rays, and the growing connections between gamma-
and x-ray astrophysics and particle physics.  At first,
some degree of skepticism may be warranted when discussing
astrophysics as a laboratory for advances in fundamental
physics.  On the other hand, there is no question that cosmology
is now in the process of making incredible strides, and should
continue to do so for the foreseeable future.  Moreover, there
are several precedents,  including a very recent and very
decisive one, for discovering new fundamental physics with
astrophysical observations or sources.  Whether these precedents
will be followed in the future will ultimately only be determined
with further vigorous cosmological experimentation.


\begin{thebibliography}{10}

\bibitem{schramm} J.-M Yang et al., ApJ 227, 697 (1979).

\bibitem{p4} D.~S.~Akerib, S.~M.~Carroll, M.~Kamionkowski and
     S.~Ritz, in {\it Proc. of the APS/DPF/DPB Summer Study on the
     Future of Particle Physics (Snowmass 2001) } ed. N.~Graf,
     arXiv:hep-ph/0201178.

\bibitem{KamSpeSug94} M. Kamionkowski, D. N. Spergel, and
     N. Sugiyama, ApJ Lett. 426, 57 (1994).

\bibitem{Miletal99} A. D. Miller et al.,
     ApJ Lett. 524, L1 (1999).

\bibitem{deBetal00} P. de Bernardis et al.,  Nature 404, 955 (2000).

\bibitem{Hanetal00} S. Hanany et al. 
     ApJ Lett. 545, L5 (2000).

\bibitem{dasiT} N. W. Halverson et al.,
     ApJ 568, 38 (2002).

\bibitem{Masetal02} B. S. Mason et al., astro-ph/0205384.

\bibitem{archeops} A. Benoit et al., astro-ph/0210306;
     astro-ph/0210305. 

\bibitem{Peretal99} S. Perlmutter et al., ApJ 517, 565 (1999).

\bibitem{Rieetal98} A. G. Riess et al., Astron. J. 116, 1009 (1998).

\bibitem{KamSpe94} E.g., M. Kamionkowski and D. N. Spergel,
     ApJ 432, 7 (1994).

\bibitem{broeils} K. G. Begeman, A. H. Broeils, and R. H. Sanders,
      MNRAS 249, 523 (1991).

\bibitem{bbn} K. A. Olive et al., ApJ 376, 51 (1991);
      S. Burles et al., Phys. Rev. Lett. 82, 4176 (1999).

\bibitem{Nbody} S. D. M. White, C. S. Frenk, and M. Davis,
      ApJ 274, L1 (1983).

\bibitem{gunn} S. Tremaine and J. E. Gunn, Phys. Rev. Lett. 42,
     407 (1979); J. Dalcanton and C. J. Hogan,
     ApJ 561, 35 (2001).

\bibitem{jkg} G. Jungman, M. Kamionkowski, and K. Griest,
     Phys. Rep. 267, 195 (1996).

\bibitem{bergstrom} L. Bergstr\"om, Rept. Prog. Phys. 63, 793 (2000).

\bibitem{axion} For reviews, see, e.g., M. S. Turner,
      Phys. Rep. 197, 67 (1990); G. G. Raffelt,
     Phys. Rep. 198, 1 (1990); L. J. Rosenberg and K. A. van
     Bibber, Phys. Rep. 325, 1 (2000).

\bibitem{unitarity} K. Griest and M. Kamionkowski,
     Phys.~Rev. Lett. 64, 615 (1990). 

\bibitem{heidelberg} M. Beck, Nucl. Phys. (Proc. Suppl.) B
     35, 150 (1994); M. Beck et al., Phys. Lett. B 336,
     141 (1994); S. P. Ahlen et al., Phys. Lett. B
     195, 603 (1987); D. O.~ Caldwell et al.,
     Phys.~Rev.~Lett. 61, 510 (1988).

\bibitem{kamiokande} M.~Mori et al.,
	Phys.~Lett. B 289, 463 (1992); M.~Mori et
	al.,  Phys.~Rev. D 48,
	5505 (1993).

\bibitem{imb} J. M.~LoSecco et al., Phys.~Lett.
	 B 188, 388 (1987).

\bibitem{baksan} M. M.~Boliev et al., Bull.\ Acad.\
       Sci.\ USSR, Phys.\ 
       Ser. 55, 126 (1991) [Izv.\ Akad.\ Nauk.\ SSSR,
       Fiz. 55, 748 (1991)]; M. M.~Boliev et al., in {\em
       TAUP 95}, proceedings of the Workshop, Toledo, Spain,
       September 17--21, 1995, ed. A.~Morales, J.~Morales
       and J. A.~Villar, [Nucl.\ Phys.\ (Proc.\ Suppl.) B
       48, 83 (1996)] (North-Holland, Amsterdam, 1996).

\bibitem{macronew} M. Ambrosio et al. (MACRO collaboration),
     Phys. Rev. D 60, 082002 (1999).

\bibitem{superK} Y. Fukuda et al., Phys. Rev. Lett. 81, 1562 (1998).

\bibitem{AMANDA} E. Andres \etal, Nucl. Phys. B
     (Proc. Suppl.) 70, 448 (1999).

\bibitem{griestsilk} K.~Griest and J.~Silk, Nature
     343, 26 (1990); L. M. Krauss, Phys.~Rev.~Lett.
     64, 999 (1990).

\bibitem{haberkane} H. E.~Haber and G. L.~Kane, Phys. Rep. 117,
     75 (1985).

\bibitem{sneutrino} T. Falk, K. A. Olive, and M. Srednicki,
     Phys. Lett. B 339, 248 (1994).

\bibitem{ellishag} J.~Ellis et al., Nucl.~Phys. B
     238, 453 (1984); K. Griest, M. Kamionkowski, and
     M. S. Turner, Phys.~Rev. D 41, 3565 (1990);
     K. A. Olive and M. Srednicki, Phys. Lett. B 230, 78 (1989);
     K. A. Olive and M. Srednicki, Nucl. Phys. B 355, 208 (1991).

\bibitem{witten} M. W.~Goodman and E.~Witten, Phys. Rev. D
     31, 3059 (1986); I. Wasserman, Phys.~Rev. D
     33, 2071 (1986); A.~Drukier, K.~Freese, and D. N. Spergel,
     Phys. Rev. D 33, 3495 (1986).

\bibitem{kim} K.~Griest, Phys. Rev. D 38, 2357 (1988);
     FERMILAB-Pub-89/139-A (E).

\bibitem{labdetectors} J. Low Temp. Phys 93 (1993);
     P. F. Smith and J. D. Lewin, Phys. Rep 187, 203 (1990).

\bibitem{SOS} J.~Silk, K. A.~Olive, and M.~Srednicki,
     Phys.~Rev.~Lett. 55, 257 (1985); K.~Freese,
     Phys. Lett. B 167, 295 (1986); L. M.~Krauss,
     K.~Freese, D. N.~Spergel, and W. H. Press,
     ApJ 299, 1001 (1985); L. M.~Krauss,
     M.~Srednicki, and F.~Wilczek, Phys.~Rev. D 33,
     2079 (1986); T.~Gaisser, G.~Steigman, and
     S.~Tilav, Phys.~Rev. D 34, 2206 (1986);
     M.~Kamionkowski, Phys.~Rev. D 44, 3021 (1991);
     F. Halzen, M. Kamionkowski, and T. Stelzer,
     Phys. Rev. D 45, 4439 (1992).

\bibitem{JunKam95} S. Ritz and D. Seckel, Nucl. Phys. B 304, 877
     (1988); G. Jungman and M. Kamionkowski, Phys. Rev. D 51,
     328 (1995).

\bibitem{pressspergel} W.~H.~Press and D.~N.~Spergel, ApJ
       296, 679 (1985); A.~Gould, ApJ 321, 571 (1987);
       A.~Gould, ApJ 388, 338 (1991).

\bibitem{IceCube} {\tt http://www.sec.wisc.edu/a3ri/icecube}.

\bibitem{dama}  R.~Bernabei et al.,
     Phys. Lett. B.480, 23 (2000).

\bibitem{cdms} R.~Abusaidi et al.,
     Phys. Rev. Lett. 84, 5699 (2000).

\bibitem{edelweiss} A. Benoit et al., astro-ph/0206271. 

\bibitem{zeplin} S. Hart, talk at  Dark
     Matter 2002, Marina del Rey, CA, Feb 2002.

\bibitem{piero} P. Ullio, M. Kamionkowski, and P. Vogel, JHEP
     0107, 044 (2001).

\bibitem{kurylov} A. Kurylov and M. Kamionkowski, in preparation.

\bibitem{modelindependent} M. Kamionkowski et al.,
     Phys. Rev. Lett. 74, 5174 (1995); M. Kamionkowski and
     K. Freese, Phys. Rev. D 55, 1771 (1997).

\bibitem{baltz} E. A. Baltz et al., Phys. Rev. D 65, 063511 (2002).

\bibitem{positrons} M. Kamionkowski and M. S. Turner,
     Phys. Rev. D 43, 1774 (1991).

\bibitem{heatpositrons} S. Coutu et al., Astropart. Phys. 11,
     429 (1999).

\bibitem{JunKam94} E.g. G. Jungman and M. Kamionkowski,
     Phys. Rev. D 49, 2316 (1994).

\bibitem{pieropbar} L. Bergstr\"om, J. Edsj\"o, and P. Ullio,
     ApJ 526, 215 (1999); P. Ullio, astro-ph/9904086.

\bibitem{berkap} L. Bergstr\"om and J. Kaplan, Astropart. Phys. 2, 261 (1994).

\bibitem{pieroZphoton} P. Ullio and L. Bergstr\"om, Phys. Rev. D 
     57, 1962 (1998); Z. Bern, P. Gondolo, and
     M. Perelstein, Phys. Lett. B 411, 86 (1997).

\bibitem{GS} P. Gondolo and J. Silk, Phys. Rev. Lett. 83, 1719
     (1999).

\bibitem{zhao} P. Ullio, H.-S. Zhao, and M. Kamionkowski,
     Phys. Rev. D 64, 043504 (2001); D. Merritt et al.,
     Phys. Rev. Lett. 88, 191301 (2002).

\bibitem{PQ} R. D. Peccei and H. R. Quinn, Phys. Rev. Lett.
     38, 1440 (1977); F. Wilczek, Phys. Rev. Lett. 40,
     279 (1978); S. Weinberg, Phys. Rev. Lett. 40, 223
     (1978).

\bibitem{gravity} M. Kamionkowski and J. March-Russell,
     Phys. Lett. B 282, 137 (1992); R. Holman et al.,
     Phys. Lett. B 282, 132 (1992); S. M. Barr and
     D. Seckel, Phys. Rev. D 46, 539 (1992);

\bibitem{solutions} R. Holman et al., Phys. Lett. B 282,
     132 (1992); N. Turok, Phys. Rev. Lett. 76, 1015
     (1996); R. Kallosh et al., Phys. Rev. D 52, 912
     (1995); E. A. Dudas, Phys. Lett. B 325, 124 (1994);
     K. S. Babu and S. M. Barr, Phys. Lett. B 300, 367 (1993). 

\bibitem{axionexperiments} S. Asztalos et al., Phys. Rev. D 64,
     092003 (2001); S. J. Asztalos et al.,
     ApJ Lett. 571, L27 (2002).

\bibitem{sikivie} P. Sikivie, Phys. Rev. Lett. 51, 1415
     (1983).

\bibitem{rydberg} I. Ogawa, S. Matsuki, and K. Yamamoto,
     Phys. Rev. D 53, 1740 (1996); S. Matsuki, I. Ogawa,
     and K. Yamamoto, Phys. Lett. B 336, 573 (1994).

\bibitem{CAST} {\tt http://nomadinfo.cern.ch/CAST}.

\bibitem{nfw} J. Navarro, C. S. Frenk, and S. D. M. White,
     ApJ 490, 493 (1997).

\bibitem{moore} B. Moore, Nature 370, 629 (1994).

\bibitem{ss} D. N. Spergel and P. J. Steinhardt,
     Phys. Rev. Lett. 84, 3760 (2000).

\bibitem{corecollapse} See, e.g., R. Dav\'e et al., ApJ 547, 574
     (2001); N. Yoshida et al., ApJ 544, L87 (2000);
     C. S. Kochanek and M. White, ApJ 543, 514 (2000).

\bibitem{mcdonald} J. McDonald, Phys. Rev. Lett. 88, 091304
     (2002); D. E. Holz and A. Zee, Phys. Lett. B 517, 239
     (2002).

\bibitem{jordi} J. Miralda-Escud\'e, ApJ 564, 60 (2002).

\bibitem{loewenstein} M. Loewenstein and R. Mushotzky,
     astro-ph/0208090.

\bibitem{KamKos99}  M. Kamionkowski and A. Kosowsky,
     Ann. Rev. Nucl. Part. Sci. 49, 77 (1999).

\bibitem{HuDod01} W. Hu and S. Dodelson,
     Ann. Rev. Astron. Astrophys. 40, 171 (2002).

\bibitem{church} S. Church, A. Jaffe, and L. Knox,
     astro-ph/0111203.

\bibitem{dasi} J. Kovac et al., astro-ph/0209478; E. M. Leitch
     et al., astro-ph/0209476.

\bibitem{map} {\tt http://www.map.gsfc.gov}

\bibitem{PLANCK} {\tt
     http://astro.estec.esa.nl/SA-general/Projects/Planck}

\bibitem{peaks}  R. Sunyaev and YaB. Zeldovich,
     Astrophys. Sp. Sci. 7, 3 (1970); P. J. E. Peebles and
     J. T. Yu, ApJ 162, 815 (1970).
 
\bibitem{Junetal96a} G. Jungman et al., Phys. Rev. Lett. 74, 5174 (1996). 

\bibitem{Junetal96b} G. Jungman et al., Phys. Rev. D 54, 1332 (1996).

\bibitem{boom} C. B. Netterfield et al., ApJ 571, 604 (571).

\bibitem{max} A. T. Lee et al., ApJ Lett. 561, L1 (2001).

\bibitem{inflation}  A. H. Guth, Phys. Rev. D 28, 347 (1981);
     A. D. Linde, Phys. Lett. B 108, 389 (1982); A. Albrecht and
     P. J. Steinhardt, Phys. Rev. Lett. 48, 1220 (1982).

\bibitem{perturbations}  A. H. Guth and S.-Y. Pi,
     Phys. Rev. Lett. 49, 1110 (1982); S. W. Hawking,
     Phys. Lett. B 115, 29 (1982); A. D. Linde, Phys. Lett. B
     116, 335 (1982); A. A. Starobinsky, Phys. Lett. B 117, 175
     (1982); J. M. Bardeen, P. J. Steinhardt, and M. S.
     Turner, Phys. Rev. D 46, 645  (1983).

\bibitem{AbbWis84} L. F. Abbott and M. Wise, Nucl. Phys. B 244,
     541 (1984).

\bibitem{KamKosSte97} M. Kamionkowski, A. Kosowsky, and
     A. Stebbins, Phys. Rev. Lett. 78, 2058 (1997).

\bibitem{SelZal97} U. Seljak and M. Zaldarriaga,
     Phys. Rev. Lett. 78, 2054 (1997).

\bibitem{KamKos98} M. Kamionkowski and A. Kosowsky, Phys. Rev. D
     67, 685 (1998).

\bibitem{JafKamWan} A. Jaffe, M. Kamionkowski, and L. Wang,
     Phys. Rev. D 61, 083501 (2000).

\bibitem{ZalSel98} M. Zaldarriaga and U. Seljak, Phys. Rev. D
     58, 023003 (1998).

\bibitem{LewChaTur02}	A. Lewis, A. Challinor, and N. Turok,
     Phys. Rev. D 65, 023505 (2002).

\bibitem{KesCooKam02} M. Kesden, A. Cooray, and M. Kamionkowski,
     Phys. Rev. Lett. 89, 011304 (2002).

\bibitem{KnoSon02} L. Knox and Y.-S. Song, Phys. Rev. Lett. 89,
     011303 (2002).

\bibitem{SelZal99} U. Seljak and M. Zaldarriaga,
     Phys. Rev. Lett. 82, 2636 (1999); M. Zaldarriaga and
     U. Seljak, Phys.Rev.D 59, 123507 (1999); U. Seljak and
     M. Zaldarriaga, Phys. Rev. D 60, 043504 (1999); W. Hu,
     Phys. Rev. D 64, 083005 (2001); W. Hu,
     ApJ Lett. 557, L79 (2001); W. Hu and T. Okamoto,
     ApJ 574, 566 (2002).

\bibitem{cozumel} M. Kamionkowski, astro-ph/0209273.

\bibitem{2dF} {\tt http://www.aao.gov.au/2df}.

\bibitem{SDSS} {\tt http://www.sdss.org}.

\bibitem{KaiWilLup00} N. Kaiser, G. Wilson, and G. A. Luppino,
     astro-ph/0003338; D. J. Bacon, A. R. Refregier, and
     R. S. Ellis, MNRAS 318, 625 (2000);
     D. M. Wittman et al., Nature 405, 143 (2000); L. van
     Waerbeke et al., Astron. Astrophys. 358, 30 (2000).

\bibitem{KamLid} M. Kamionkowski and A. R. Liddle,
     Phys. Rev. Lett. 84, 4525 (2000).

\bibitem{threept} F. Bernardeau et al., Phys. Rep. 367, 1 (2002).

\bibitem{verde} L. Verde et al., MNRAS 325, 412 (2001).

\bibitem{verdetwo} L. Verde et al., MNRAS 321, L7 (2001).

\bibitem{Carroll:2000fy} S. Carroll, Living Rev. Rel. 4, 1 (2001).

\bibitem{Caletal98} R. R. Caldwell, R. Dave, and
     P. J. Steinhardt, Phys. Rev. Lett. 80, 1582 (1998);
     B. Ratra and P. J. E. Peebles, Phys. Rev. D 37, 3406
     (1998); K. Coble, S. Dodelson, and J. A. Frieman,
     Phys. Rev. D 55, 1851 (1997); M. S. Turner and M. White,
     Phys. Rev. D 56, 4439 (1997).

\bibitem{BoyCalKam01} L. A. Boyle, R. R. Caldwell, and
     M. Kamionkowski, Phys. Lett. B 545, 17 (2002); J.-A. Gu and
     W.-Y. P. Hwang, astro-ph/0105099.

\bibitem{caldwell} R. R. Caldwell, Phys. Lett. B 545, 23
     (2002).

\bibitem{DMT} J. A. Tyson and D. Wittman, astro-ph/0005381.

\bibitem{WFHRI} N. Kaiser, J. L. Tonry, and G. A. Luppino,
     P. Astron. Soc. Pac. 112, 768 (2000).

\bibitem{SNAP} {\tt http://snap.lbl.gov}.

\bibitem{nevinone} N. N. Weinberg and M. Kamionkowski, 
     astro-ph/0203061.

\bibitem{nevintwo} N. N. Weinberg and M. Kamionkowski, 
     astro-ph/02010134.

\bibitem{Carroll} S. M. Carroll, Phys. Rev. Lett. 81, 3067 (1998).

\bibitem{LueWanKam98} A. Lue, L, Wang, and M. Kamionkowski,
     Phys. Rev. Lett. 83, 1506 (1999).

\end{thebibliography}
\end{document}